\documentclass[twocolumn,showpacs,amsmath,amssymb,aps,prx,floatfix,longbibliography, 10pt]{revtex4-2}

\usepackage[latin1]{inputenc}
\usepackage[american]{babel}
\usepackage[T1]{fontenc}
\usepackage{graphicx}
\usepackage{amsfonts}
\usepackage{amsmath}
\usepackage{bbold}
\usepackage{mathrsfs}
\usepackage{hyperref}
\usepackage{epstopdf}
\usepackage{bm, epsfig}

\usepackage[tbtags]{mathtools}
\usepackage{afterpage}

\DeclarePairedDelimiterX\braket[2]{\langle}{\rangle}{#1 \delimsize\vert #2}
\hypersetup{colorlinks=true, urlcolor=blue, citecolor=blue, filecolor=blue, linkcolor=blue}
\newcommand*{\diff}{\mathop{}\!\mathrm{d}}

\newcommand*{\Imm}{\mathop{}\!\mathbf{Im}}

\newcommand{\uimm}{\mathrm{i}}
\newcommand{\eu}{\mathrm{e}}

\newcommand{\Tr}{\mathrm{Tr}}
\newcommand{\daga}{^{\dagger}}

\newcommand*{\sinc}{\mathop{}\!\mathrm{sinc}}

\newcommand*{\szero}{S\textsubscript{0} }
\newcommand*{\sone}{S\textsubscript{1} }
\newcommand*{\stwo}{S\textsubscript{2} }
\newcommand*{\tc}{TRUECARS }
\newcommand*{\ctc}{c-TRUECARS }
\newcommand*{\stc}{s-TRUECARS }

\begin{document}

\title{High temporal and spectral resolution of stimulated x-ray Raman signals with stochastic free-electron-laser pulses}

\author{Stefano~M.~Cavaletto}
\affiliation{Department of Chemistry and Department of Physics \& Astronomy, University of California, Irvine, CA 92697, USA}
\author{Daniel~Keefer}
\affiliation{Department of Chemistry and Department of Physics \& Astronomy, University of California, Irvine, CA 92697, USA}
\author{Shaul~Mukamel}
\email[Email: ]{smukamel@uci.edu}
\affiliation{Department of Chemistry and Department of Physics \& Astronomy, University of California, Irvine, CA 92697, USA}
\date{\today}

\begin{abstract}

The chaotic nature of x-ray free-electron-laser pulses is a major bottleneck that has limited the joint temporal and spectral resolution of spectroscopic measurements. We show how to use the stochastic x-ray field statistics to overcome this difficulty by correlation signals averaged over independent pulse realizations. No control is required over the spectral phase of the pulse, enabling immediate application with existing, noisy x-ray free-electron-laser pulses. The proposed stimulated Raman technique provides the broad observation bandwidth and high time--frequency resolution needed for the observation of elementary molecular events. A model is used to simulate chaotic free-electron-laser pulses and calculate their correlation properties. The resulting joint temporal/spectral resolution is exemplified for a molecular model system with time-dependent frequencies and for the RNA base Uracil passing through a conical intersection. Ultrafast coherences, which represent a direct signature of the nonadiabatic dynamics, are resolved. The detail and depth of physical information accessed by the proposed stochastic signal are virtually identical to those obtained by phase-controlled pulses.
\end{abstract}


\maketitle

\section{Introduction}

Recent advances in the generation of sub-femtosecond extreme-ultraviolet (XUV) and x-ray pulses are enabling the control of electron dynamics on their natural time scales \cite{krausz2009attosecond, pellegrini2016physics, duris2020tunable,maroju2020attosecond}. This is essential for the direct manipulation of the ensuing electronic and nuclear dynamics and for the control of chemical reactions with light, with broad applications to photochemistry and photobiology \cite{kraus2015measurement, nisoli2017attosecond, worner2017charge}.

Free-electron lasers (FELs) provide intense pulses at frequencies ranging from the XUV to the hard-x-ray domain \cite{pellegrini2016physics} suitable for nonlinear x-ray spectroscopy \cite{bennett2016multidimensional}. While XUV seeded FELs offer stable coherent XUV pulses \cite{allaria2012highly} with the possibility of pulse shaping \cite{gauthier2015spectrotemporal} and control \cite{prince2016coherent}, soft- and hard-x-ray FELs based on the self-amplified spontaneous emission (SASE) mechanism \cite{bonifacio1984collective} provide stochastic pulses with limited longitudinal coherence, and noisy spikes in their temporal and spectral profiles. Stimulated x-ray Raman scattering, a fundamental building block of nonlinear spectroscopy \cite{rohringer2019x}, was recently demonstrated using hard-x-ray FEL pulses \cite{weninger2013stimulatedPRL}, but future multidimensional nonlinear x-ray spectroscopy protocols \cite{biggs2013watching, healion2012entangled, zhang2014monitoring} require coherent and reproducible pulses. Self- or laser-seeding methods have been implemented to improve the coherence of hard-x-ray FEL pulses \cite{amann2012demonstration}. Novel techniques have demonstrated high-intensity few-femtosecond pulses \cite{marinelli2015high} with reduced intensity spikes \cite{lutman2018high}, but with an underlying SASE structure which renders them not reproducible from shot to shot.

\begin{figure}[b]
\centering
\includegraphics[width=\linewidth]{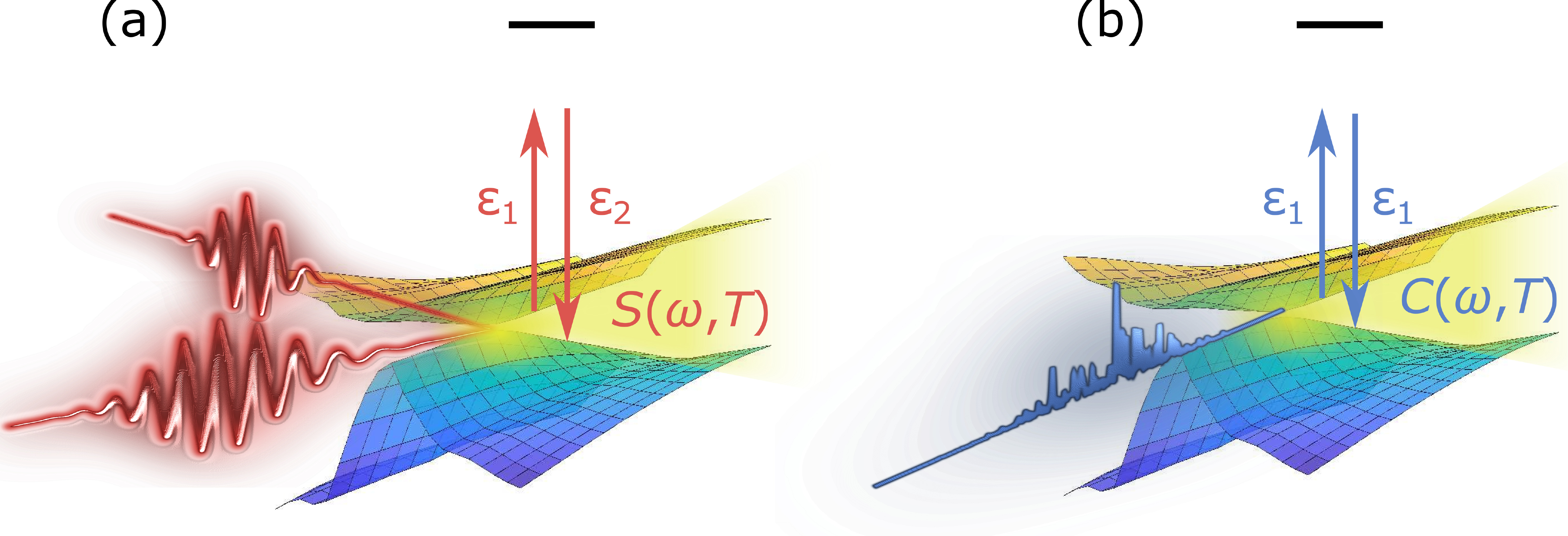}
\caption{Off-resonant stimulated x-ray Raman signal (TRUECARS) between two molecular potential-energy surfaces via an off-resonant core state. (a) TRUECARS with hybrid broadband ($\mathcal{E}_2$) and narrowband ($\mathcal{E}_1$) coherent x-ray pulses (red), which requires control over their CEPs; and (b) s-TRUECARS with one stochastic x-ray FEL pulse ($\mathcal{E}_1$, blue) with no phase control.}
\label{fig:TRUECARS}
\end{figure}

Recently, transient redistribution of ultrafast electronic coherences in attosecond Raman signals (TRUECARS) \cite{kowalewski2015catching} was proposed as a suitable technique to achieve the demanding time--frequency resolution necessary to detect ultrafast nonadiabatic molecular processes, such as at conical intersections (CoIns) of electronic states \cite{worth2004beyond,domcke2011conical}. CoIns are ubiquitous in molecules, playing an essential role in virtually all photochemical processes. However, their direct experimental observation, i.e., via signals whose appearance can be exclusively attributed to CoIns, is a challenging task due to the requirements on joint temporal and spectral resolution. Several approaches have addressed this issue \cite{polli2010conical, oliver2014correlating, mcfarland2014ultrafast}. In TRUECARS, as shown in Fig.~\ref{fig:TRUECARS}(a), two coherent x-ray pulses induce an off-resonant stimulated x-ray Raman process between the valence electronic states involved in the CoIn. By varying the pulses' arrival time, a time-resolved measurement of this signal allows direct, background-free access to the nonadiabatic dynamics of the molecular coherences. This differs from other approaches that possess additional contributions from the populations \cite{kobayashi2019direct, timmers2019disentangling} or use strong fields \cite{sussman2006dynamic, corrales2014control}. The TRUECARS technique, however, assumes reproducible coherent pulses with control over their carrier-envelope phases (CEPs). This hinders its application with existing x-ray technology.

Here, we show how the technique can be implemented with stochastic x-ray FEL pulses. The proposed stochastic (s-TRUECARS) technique, displayed in Fig.~\ref{fig:TRUECARS}(b), does not require control over the pulse phase. By averaging over independent pulse realizations, s-TRUECARS takes advantage of correlations between the spectral components of the field, providing joint temporal and spectral resolutions comparable to TRUECARS with phase-controlled pulses (c-TRUECARS). Correlation techniques have been investigated in recent theoretical and experimental studies with stochastic optical lasers \cite{tollerud2019femtosecond,osipov2019time} and x-ray FEL pulses \cite{kimberg2016stochastic, gorobtsov2018seeded, asban2019frequency, kayser2019core, driver2020attosecond}. We use a model of stochastic x-ray FEL pulses that describes their amplitude and phase fluctuations. The s-TRUECARS performance is illustrated for a model system with two electronic states with time-dependent frequencies and further applied to the RNA base Uracil undergoing a CoIn. The methods presented here, based on correlation functions of stochastic FEL fields, can be straightforwardly extended to other time-resolved nonlinear x-ray signals, including multidimensional nonlinear spectroscopy.

The paper is structured as follows. In Sec.~\ref{Sec:PCM}, we introduce the model of stochastic FEL pulses and calculate the relevant multi-point field correlation functions. These are the crucial quantities determining the average properties of the signal. The properties of c-TRUECARS implemented with coherent phase-controlled pulses are presented in Sec.~\ref{Sec:TRUECARS} and demonstrated for a model with time-dependent frequencies. Section~\ref{Sec:s-TRUECARS} presents the s-TRUECARS signal, illustrated for a time-dependent-frequency model (Sec.~\ref{Sec:s-TRUECARS-TDF}) and applied to Uracil (Sec.~\ref{Sec:s-TRUECARS-CoIn}). Finally, in Sec.~\ref{Sec:Conclusions}, we discuss future extensions of the methods used here to multidimensional nonlinear x-ray spectroscopies.

\section{Modeling of stochastic x-ray free-electron-laser pulses}
\label{Sec:PCM}
SASE FEL pulses arise from the self-amplification of the photons spontaneously emitted by an electron beam in a linear accelerator \cite{pellegrini2016physics}. The interaction between the electron beam and the initially emitted photons creates electron bunches, which emit intense bursts of in-phase x~rays. Due to the noisy nature of the spontaneously emitted photons involved in the process, FEL pulses feature chaotic envelopes and a limited longitudinal coherence. The temporal envelope of an FEL pulse consists of a series of short spikes spanning the overall duration of the pulse. Each spike has an average duration given by the pulse coherence time. The spectrum of an FEL pulse has a similarly spiky structure, with several peaks within its bandwidth (see Fig.~\ref{fig:PulseProfileMaxVariance}).

Early optical-laser experiments were also performed with the chaotic pulses available at the time, and simulation techniques were developed to model their properties \cite{vannucci1980computer}. These methods have long been utilized to model experiments at x-ray FELs \cite{rohringer2007x, pfeifer2010partial, cavaletto2012resonance, weninger2013stimulatedPRA, giri2020purifying, lyu2020narrow}. Pfeifer~\textit{et~al.}\ showed that, by using a model starting from random spectral phases, one can simulate chaotic pulses with the correct statistical properties of SASE FEL pulses, including their time and frequency spiky profiles and their energy distribution \cite{pfeifer2010partial}. Below, we briefly outline the model and present the key pulse properties, in particular the associated two- and four-point correlation functions of the field. These are then used in Sec.~\ref{Sec:s-TRUECARS} to study the s-TRUECARS signal.

\subsection{Stochastic model for pulse intensity and phase fluctuations}
\label{Sec:pulses}

The envelope of the stochastic pulse is represented by
\begin{equation}
E(t) = 2\pi\,f(t)\,u(t),
\label{eq:stochpulset}
\end{equation}
with the stochastic term $f(t)$ and the temporal gating function $u(t)$ determining the pulse spectral bandwidth and the time duration, respectively. The pulse intensity is given by $|A\,E(t)|^2/(8\pi\alpha)$, where $A$ is a prefactor ensuring the correct peak intensity, and $\alpha$ is the fine-structure constant. Atomic units are used throughout unless otherwise stated.

The function 
\begin{equation}
f(t) = \int\frac{ \diff\omega}{2\pi}\,\tilde{g}(\omega)\,\eu^{\uimm\varphi(\omega)}\,\eu^{-\uimm\omega t}
\end{equation}
is obtained via the Fourier transform of the complex function 
$\tilde{f}(\omega) = \tilde{g}(\omega)\,\eu^{\uimm\varphi(\omega)}$, with a broadband real envelope $\tilde{g}(\omega)$ and a stochastic phase $\varphi(\omega)$. $\tilde{g}(\omega)$ sets the pulse bandwidth, while the phase $\varphi(\omega)$ is obtained by interpolating a set of independent random variables $\varphi_k$, corresponding to a discrete grid of frequencies $\omega_k = k\varLambda$, where $\varLambda$ is the sampling frequency. Each $\varphi_k$ is a uniformly distributed stochastic phase (UDSP) varying in the interval $[-a,\,a]$ with probability density function
\begin{equation}
P(\varphi_k) = \left\{
\begin{aligned}
&\frac{1}{2a}, & &\text{if $-a\leq\varphi_k\leq a$},\\
&0,  &&\text{otherwise}.
\end{aligned}
\right.
\label{eq:stochphase}
\end{equation}

The temporal envelope $f(t)$ features stochastic fluctuations in its intensity and phase, with a long overall duration given by $1/\varLambda$. Multiplying $f(t)$ by the temporal gating function $u(t)$, whose duration $\tau$ is much shorter than $1/\varLambda$ but still significantly longer than the average duration of the spikes in $f(t)$, ensures that the complex envelope $E(t)$ in Eq.~(\ref{eq:stochpulset}) has a finite duration $\tau$ [see, e.g., Fig.~\ref{fig:PulseProfileMaxVariance}(a)]. As a result of the gating function $u(t)$, the spectral envelope of the gated pulse is given by
\begin{equation}
\begin{aligned}
\tilde{E}(\omega) = \,&\int \diff t\,E(t)\,\eu^{\uimm\omega t} = \int\diff\omega'\,\tilde{f}(\omega')\,\tilde{u}(\omega - \omega')\\
\approx\,& \tilde{g}(\omega)\int\diff\omega'\,\eu^{\uimm\varphi(\omega')}\,\tilde{u}(\omega - \omega'),
\end{aligned}
\label{eq:stochpulse}
\end{equation}
with stochastic fluctuations both in its spectral intensity and phase [see, e.g., Fig.~\ref{fig:PulseProfileMaxVariance}(b)] in agreement with the spiky spectral features of FEL pulses \cite{pellegrini2016physics}. The convolution in Eq.~(\ref{eq:stochpulse}) shows that $\tilde{u}(\omega)$ acts as a spectral gating function over the rapid oscillations of $\eu^{\uimm\varphi(\omega)}$. This has two important consequences: first, $\tilde{E}(\omega)$ has intensity fluctuations; second, its phase is not given by $\varphi(\omega)$. Both the amplitude and the phase of $\tilde{E}(\omega)$ vary on the broader frequency scale set by the width of $\tilde{u}(\omega)$.

We assume Gaussian envelopes
\begin{equation}
\tilde{g}(\omega) = \eu^{-\tfrac{\omega^2}{2\sigma^2}},
\end{equation}
\begin{equation}
u(t) = \frac{1}{\sqrt{2\pi}}\,\eu^{-\tfrac{t^2}{2\tau^2}},
\end{equation}
with bandwidth $\sigma$, pulse duration $\tau$, and Fourier transforms
\begin{equation}
g(t) = \int\frac{ \diff\omega}{2\pi}\,\eu^{-\tfrac{\omega^2}{2\sigma^2}}\,\eu^{-\uimm\omega t} = \frac{\sigma}{\sqrt{2\pi}}\,\eu^{-\tfrac{\sigma^2t^2}{2}}
\end{equation}
and 
\begin{equation}
\tilde{u}(\omega) = \int \diff t\,\frac{1}{\sqrt{2\pi}}\,\eu^{-\tfrac{t^2}{2\tau^2}}\,\eu^{\uimm\omega t} =\tau\,\eu^{-\tfrac{\omega^2\tau^2}{2}},
\end{equation}
respectively. We require that $\varLambda \ll 1/\tau$, in order to reproduce the spikes in the frequency envelope of the pulse, as observed experimentally. We set the additional condition $1/\tau \ll \sigma$, ensuring that the width of these spikes is narrower than the overall pulse bandwidth.

%

The outcome of nonlinear spectroscopy experiments with stochastic pulses depends on $n$-point field correlation functions $F_n(\omega_1,\omega_2,\ldots,\omega_n) = \langle\tilde{E}_1(\omega_1)\tilde{E}_2(\omega_2)\cdots \tilde{E}_n(\omega_n)\rangle$, where $\langle\cdots\rangle$ denotes the ensemble average over independent realizations. The two- and four-point correlation functions of the field $\tilde{E}(\omega)$ for our UDSP model [Eq.~(\ref{eq:stochpulse})] are given in Eqs.~(\ref{eq:F2uniformstat}) and (\ref{eq:F4uniformstat}) in Appendix~\ref{App:uniformdist}. These correlation functions are the key quantities we will use in Sec.~\ref{Sec:s-TRUECARS} to calculate the s-TRUECARS signal.

Machine drifts in the electron-bunch energy at FELs cause a shot-to-shot jitter in the central frequency $\omega_{\mathrm{X}}$ of the resulting pulse $\mathcal{E}(t) = E(t)\,\eu^{-\uimm\omega_{\mathrm{X}}t}$, which can affect the resolution of measurable absorption spectra \cite{kimberg2016stochastic}. As we show in Appendix~\ref{App:jitter}, including this energy jitter in the stochastic-pulse envelope of Eq.~(\ref{eq:stochpulse}) leads to a shot-to-shot shift in the central frequency of the envelope function $\tilde{g}(\omega)$ appearing in the two- and four-point correlation functions of the field. In the following, we assume broadband stochastic pulses with bandwidths larger than the frequency jitter caused by machine drifts, such that the shot-to-shot change in Eqs.~(\ref{eq:F2uniformstat}) and (\ref{eq:F4uniformstat}) and in the associated s-TRUECARS signal can be safely neglected.

The field measured by a detector with finite resolution $Q$ is given by an additional convolution between $\tilde{E}(\omega)$ from Eq.~(\ref{eq:stochpulse}) and the detector's response function $\tilde{q}_Q(\omega)$. X-ray detectors planned for resonant inelastic x-ray scattering (RIXS) experiments have a frequency resolution ranging from $10\,\mathrm{meV}$ at hard-x-ray frequencies to $30\,\mathrm{meV}$ for soft x~rays. As long as $Q$ is narrower than the width $1/\tau$ of the gating function $\tilde{u}(\omega)$, the effect of the finite detector resolution can be safely neglected. In our calculations, we assume pulses satisfying this condition.

\subsection{Pulse properties}

\begin{figure}[b]
\centering
\includegraphics[width=0.75\linewidth]{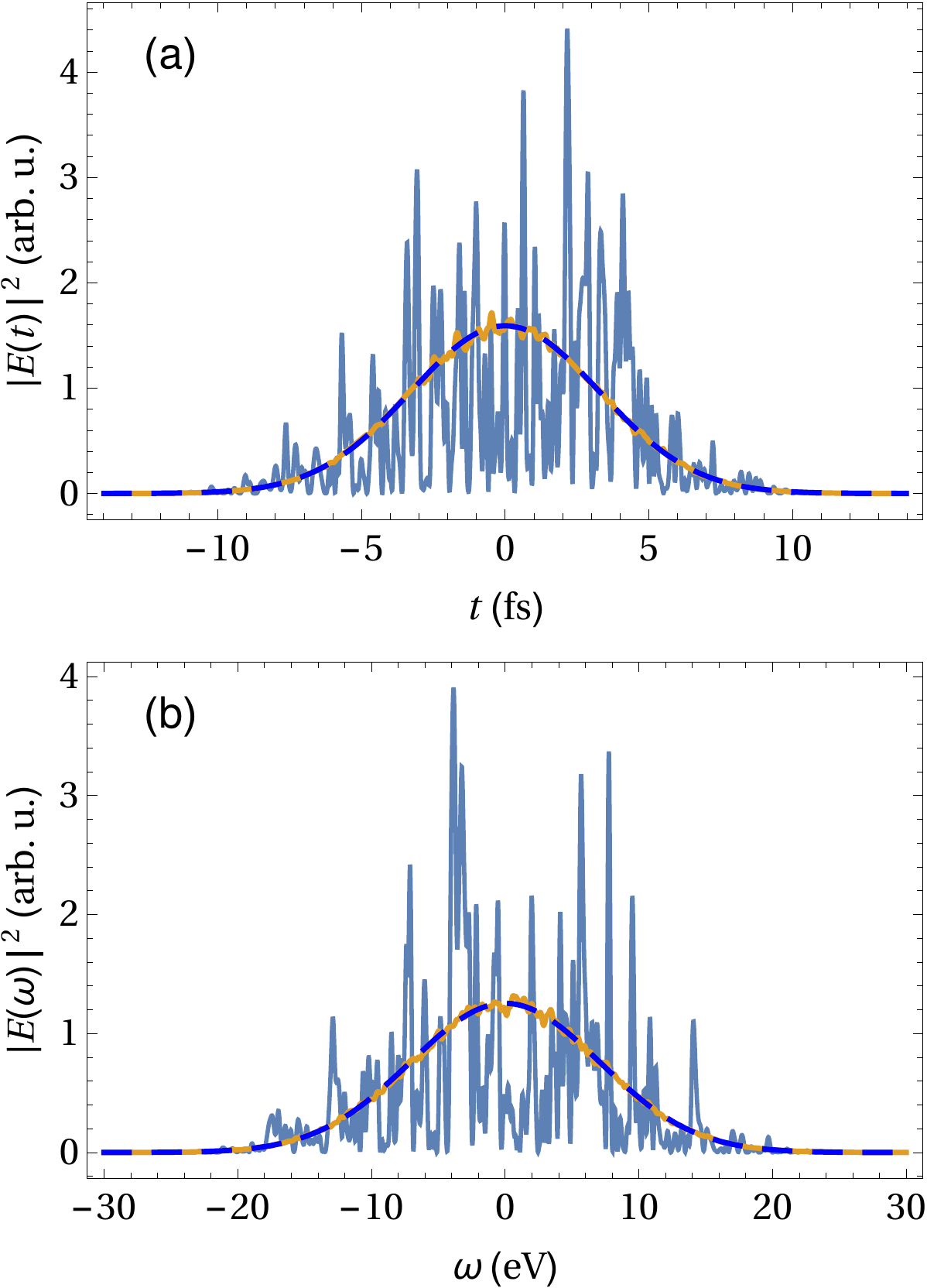}
\caption{Intensity profiles of a stochastic UDSP pulse with $a=\pi$, $\varLambda = 5\,\mathrm{meV}$, $\tau = 4.65\,\mathrm{fs}$ ($1/\tau = 0.14\,\mathrm{eV}$), and $\sigma = 10\,\mathrm{eV}$. The (a) temporal and (b) spectral intensity profiles are shown. The blue continuous curves display one realization of the stochastic pulse from Eqs.~(\ref{eq:stochpulset}) and (\ref{eq:stochpulse}). The yellow continuous curves exhibit the mean profiles of $|E(t)|^2$ and $|\tilde{E}(\omega)|^2$, obtained by averaging over 1,000 independent realizations. The blue dashed curves show the average intensities calculated from Eqs.~(\ref{eq:meantemint}) and (\ref{eq:meanspint}).}
\label{fig:PulseProfileMaxVariance}
\end{figure}

\begin{figure}[b]
\centering
\includegraphics[width=0.85\linewidth]{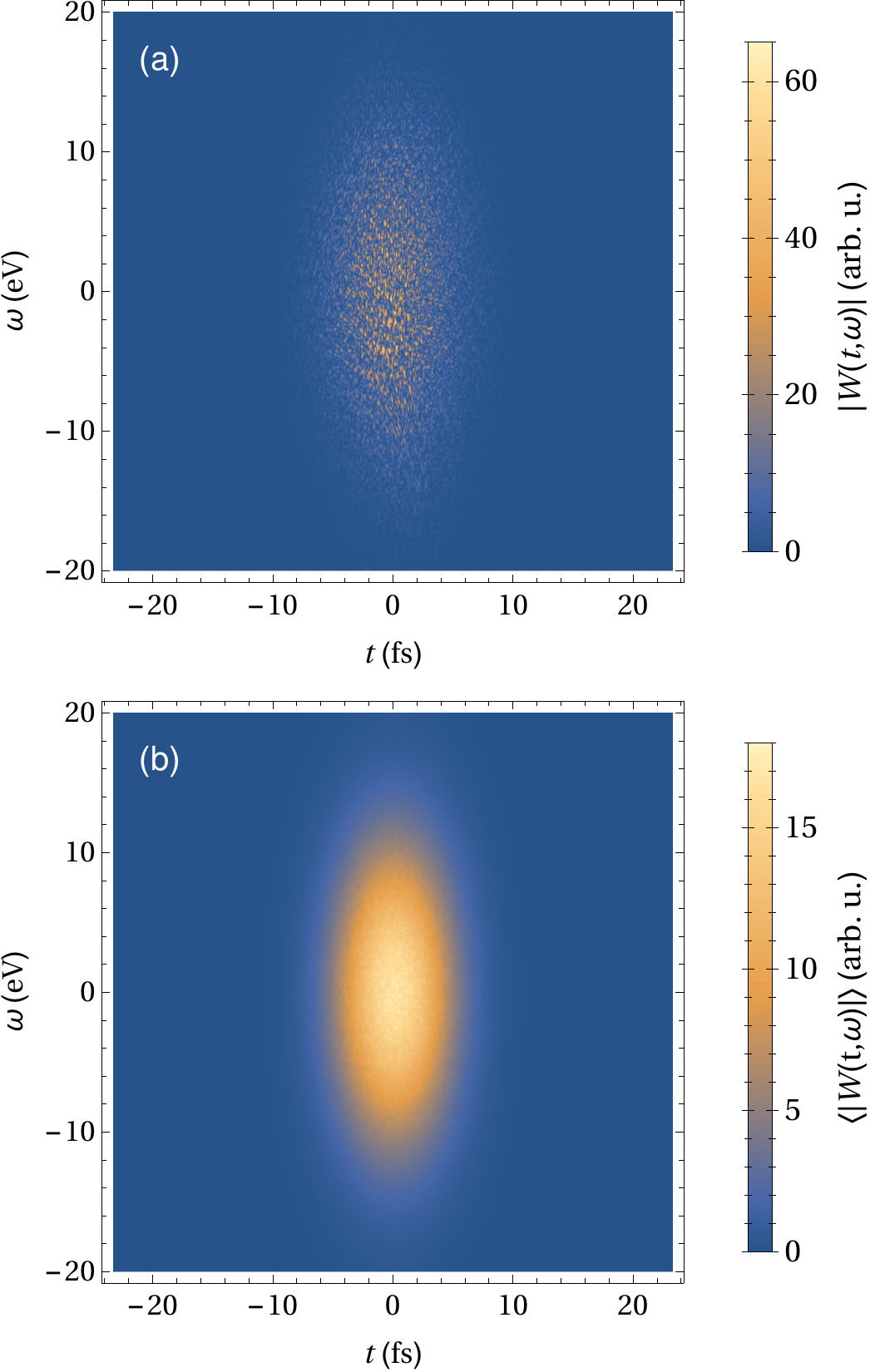}
\caption{Wigner spectrogram [Eq.~(\ref{eq:Wigner})] for the same pulse parameters as in Fig.~\ref{fig:PulseProfileMaxVariance}. (a) Modulus of the Wigner spectrogram of a single stochastic pulse realization and (b) mean modulus of the Wigner spectrogram averaged over 500 pulse realizations.}
\label{fig:WignerMaxVariance}
\end{figure}

\begin{figure}[b]
\centering
\includegraphics[width=0.75\linewidth]{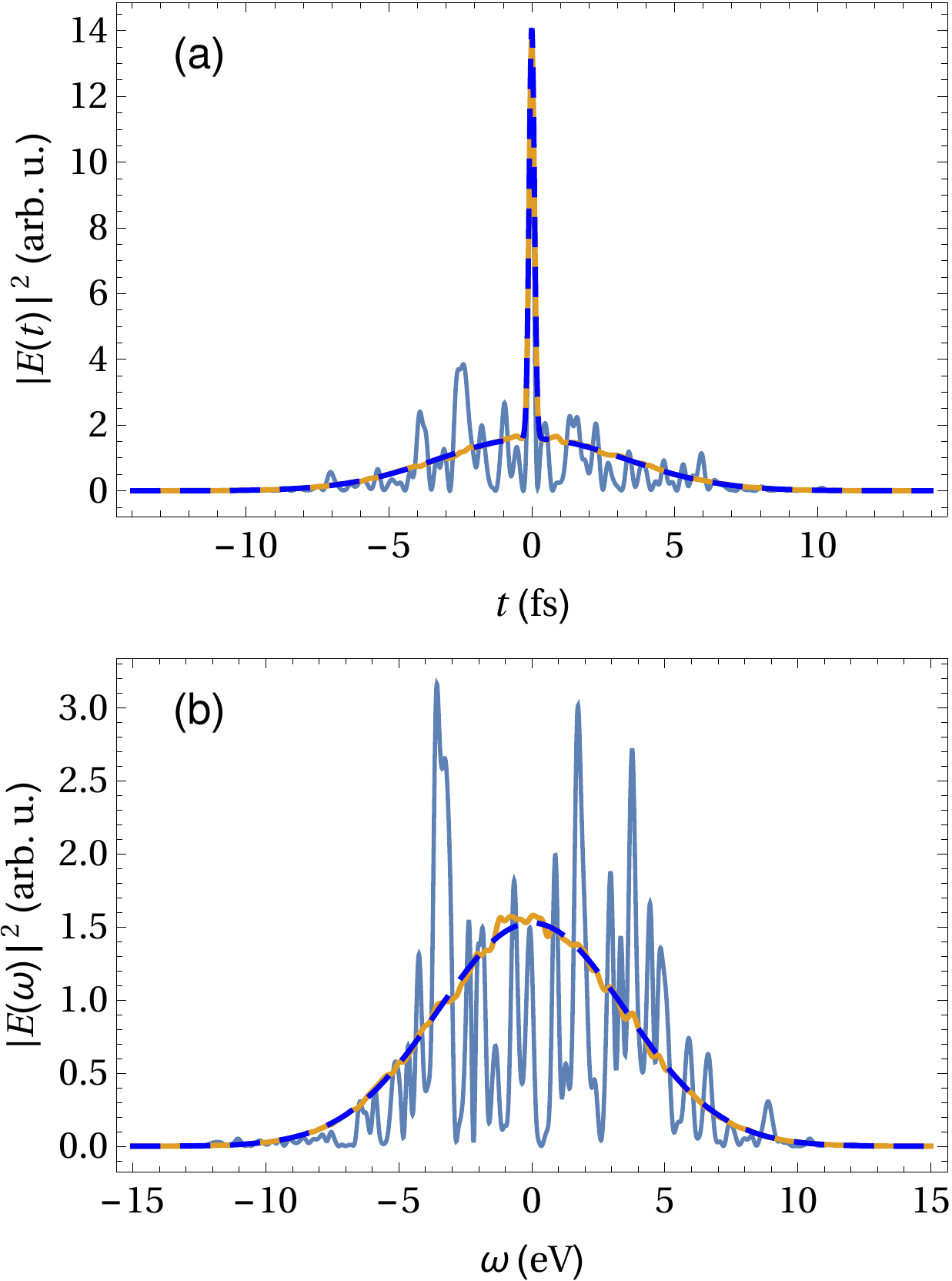}
\caption{Intensity profiles of a stochastic UDSP pulse with $a=3$, $\varLambda = 5\,\mathrm{meV}$, $\tau = 4.65\,\mathrm{fs}$ ($1/\tau = 0.14\,\mathrm{eV}$), and $\sigma = 5\,\mathrm{eV}$. The (a) temporal and (b) spectral intensity profiles are shown. The blue continuous, yellow continuous, and blue dashed curves have the same meaning as in Fig.~\ref{fig:PulseProfileMaxVariance}.}
\label{fig:PulseProfileSmallVariance}
\end{figure}

The spiky temporal and spectral profiles of stochastic UDSP pulses are shown in Figs.~\ref{fig:PulseProfileMaxVariance}, \ref{fig:WignerMaxVariance}, and \ref{fig:PulseProfileSmallVariance} for different values of the parameter $a$. The ensemble-averaged temporal and spectral intensity profiles can be calculated in terms of $F_2(\omega_1,\omega_2)$ and are given by
\begin{equation}
\begin{aligned}
\langle|E(t)|^2 \rangle  \,&= \int\frac{\diff\omega_1}{2\pi} \int\frac{\diff\omega_2}{2\pi}\,F_2(\omega_1, \omega_2)\,\eu^{-\uimm(\omega_2-\omega_1)t} \\
&\,=2\pi\,s^2(a)\, |g(t)|^2+ \left[1-s^2(a)\right]\varLambda\sqrt{\pi}\,\sigma \,|u(t)|^2
\end{aligned}
\label{eq:meantemint}
\end{equation}
and
\begin{equation}
\begin{aligned}
\langle|\tilde{E}(\omega)|^2 \rangle \,&= F_2(\omega,\omega) \\
&= \left\{2\pi\,s^2(a) + \left[1-s^2(a)\right]
\varLambda\sqrt{\pi}\tau\right\}\,|\tilde{g}(\omega)|^2,
\end{aligned}
\label{eq:meanspint}
\end{equation}
with $s(a) = \sinc(a) = \sin(a)/a$.

In Fig.~\ref{fig:PulseProfileMaxVariance}, we display stochastic UDSP pulses with $a = \pi$. This case was shown to reproduce the statistical properties of experimental SASE FEL pulses, i.e., their energy distribution and time and frequency spiky profiles \cite{pfeifer2010partial}. The temporal intensity profile of a single pulse from Eq.~(\ref{eq:stochpulset}) is shown by the blue continuous curve in Fig.~\ref{fig:PulseProfileMaxVariance}(a). Its duration $\tau$ is associated with the time envelope $u(t)$, with several short spikes of average duration determined by the inverse $1/\sigma$ of the pulse bandwidth. The blue continuous curve in Fig.~\ref{fig:PulseProfileMaxVariance}(b) represents the spectral intensity of the same stochastic pulse from Eq.~(\ref{eq:stochpulse}). It has an overall width $\sigma$ given by the frequency envelope $\tilde{g}(\omega)$, with spikes of average width $1/\tau$ owing to the finite pulse duration. In both panels, the yellow curves, obtained by averaging over independent realizations of the stochastic pulse, are in very good agreement with the mean temporal and spectral intensities [Eqs.~(\ref{eq:meantemint}) and (\ref{eq:meanspint})], shown by the blue dashed curves. For $a = \pi$ and $s(a) = 0$, these reduce to $\langle|E(t)|^2 \rangle =\varLambda\sqrt{\pi}\,\sigma \,|u(t)|^2$ and $\langle|\tilde{E}(\omega)|^2 \rangle =\varLambda\sqrt{\pi}\tau\,|\tilde{g}(\omega)|^2$, respectively, and are independently determined by the time and frequency envelopes $|u(t)|^2$ and $|\tilde{g}(\omega)|^2$. 

To illustrate the time--frequency pulse profiles of this model, we examine the Wigner spectrogram of the pulse envelope $E(t)$
\begin{equation}
W(t,\omega) = \int\diff \tau E^*\left(t + \frac{\tau}{2}\right)\,E\left(t - \frac{\tau}{2}\right)\,\eu^{-\uimm\omega\tau}.
\label{eq:Wigner}
\end{equation}
Figure~\ref{fig:WignerMaxVariance}(a) displays the modulus $|W(t,\omega)|$ for a single stochastic pulse, while the expectation value of $|W(t,\omega)|$, obtained by averaging over several independent realizations, is shown in Fig.~\ref{fig:WignerMaxVariance}(b). The temporal and spectral widths of the pulse are determined by $\tau$ and $\sigma$, respectively, and their product is larger than the Fourier uncertainty minimum $\tau\sigma>1$.

We next consider UDSP pulses with $a<\pi$. Figure~\ref{fig:PulseProfileSmallVariance} shows simulation results for $a = 3$. The temporal and spectral profiles exhibit similar patterns to Fig.~\ref{fig:PulseProfileMaxVariance}. However, a central peak now emerges in the temporal intensity profile of the pulse, as apparent in Fig.~\ref{fig:PulseProfileSmallVariance}(a). The blue continuous curve shows a single stochastic pulse, featuring a central peak surrounded by a noisy background. This central peak survives in the average temporal intensity of the pulse, depicted by the yellow continuous curve, and provides a clearly defined central time. This agrees with the average time intensity in Eq.~(\ref{eq:meantemint}) for $s(a)\neq 0$: two contributions are present, respectively proportional to $|g(t)|^2$ and $|u(t)|^2$ and both recognizable in the blue dashed curve in Fig.~\ref{fig:PulseProfileSmallVariance}(a). The central peak of UDSP pulses with $a<\pi$ is reminiscent of the properties observed in phase-gate shaped optical pulses \cite{oron2002narrow} used for the control of resonant Raman signals of vibrational states.

UDSP pulses with $a<\pi$ have not been previously used to model chaotic SASE FEL pulses. However, they can be realized by pulse-shaping capabilities at FELs \cite{gauthier2015spectrotemporal}, which can generate stochastic FEL pulses with engineered amplitudes and phases. In addition, the correlation properties of UDSP pulses with $a<\pi$ can be achieved via an alternative stochastic-pulse scheme presented in Appendix~\ref{Appendix:C+S}. In this case, the stochastic pulse in Eq.~(\ref{eq:sumpulse}) is given by the sum of a short broadband pulse and a stochastic UDSP FEL pulse with $a = \pi$, without requiring any shaping or control of the pulse phase. The associated two- and four-point correlation functions [Eqs.~(\ref{eq:F2sumpulse}) and (\ref{eq:F4sumpulse})] exhibit a structure analogous to UDSP pulses with $a<\pi$ [Eqs.~(\ref{eq:F2uniformstat}) and (\ref{eq:F4uniformstat})], which translates into identical s-TRUECARS signals for these two stochastic-pulse models. This will be further discussed in Sec.~\ref{Sec:s-TRUECARS}

\section{TRUECARS with hybrid broad-/narrowband coherent pulses}
\label{Sec:TRUECARS}
Stimulated Raman spectroscopy has been widely employed with near-infrared and optical fields to monitor vibrational dynamics of molecules. Impulsive stimulated Raman spectroscopy uses off-resonant femtosecond pulses to induce a Raman process between two vibrational states \cite{dhar1994time, kukura2007femtosecond}. Augmenting the broadband pulse with an additional picosecond pulse was shown to improve the joint time--frequency resolution \cite{kukura2005structural, mukamel2011communication,dorfman2013time}.

\begin{figure}[b]
\centering
\includegraphics[width=\linewidth]{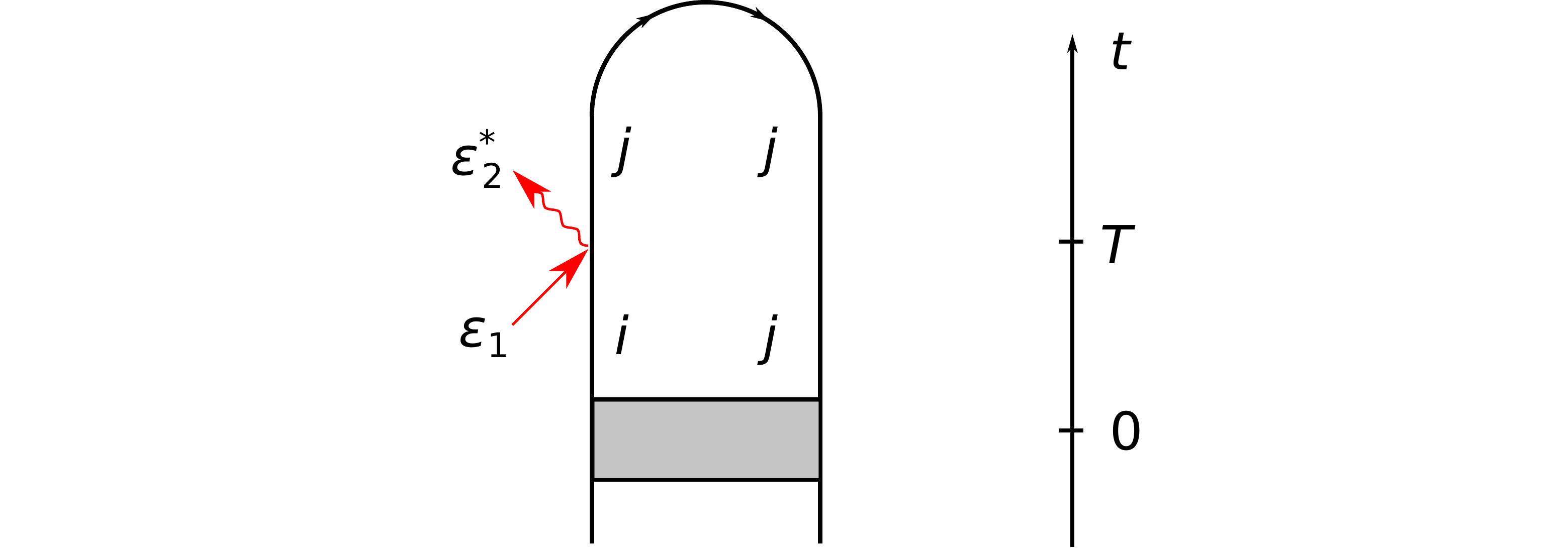}
\caption{Loop diagram of the off-resonant stimulated Raman signal TRUECARS. The red arrows represent the fields $\mathcal{E}_1$, exciting the system, and $\mathcal{E}^*_2$, stimulating the emission of the signal photon.}
\label{fig:Feynman}
\end{figure}

Intense coherent HHG- and FEL-based XUV and x-ray pulses can induce stimulated Raman excitations for the study of electronic valence-state dynamics \cite{bennett2016multidimensional}. Resonant stimulated Raman scattering has been recently demonstrated in neon with a hard-x-ray FEL pulse \cite{weninger2013stimulatedPRL}. Inspired by stimulated Raman spectroscopy of vibrational states with a femtosecond and a picosecond optical pulse, the TRUECARS signal was proposed to monitor nonadiabatic molecular processes and the associated fast electronic dynamics via the combination of an attosecond and a femtosecond pulse \cite{kowalewski2015catching, kowalewski2017simulating}. In contrast to conventional stimulated Raman spectroscopy, which is a quartic Raman signal in which each pulse interacts twice with the system, TRUECARS is linear in both pulse amplitudes, as shown in Fig.~\ref{fig:Feynman}. The signal has no contributions from level populations and can thus directly access the evolution of electronic coherences in a background-free manner, rendering it a direct signature of CoIns. The TRUECARS signal implemented with two coherent pulses (c-TRUECARS) requires control over their CEPs. This limits its implementation with existing intense stochastic FEL pulses.

In the following, we summarize the key features of the c-TRUECARS signal, showing how two pulses of different bandwidth can provide independent control over the observation bandwidth of the technique and its time--frequency resolution. This sets the stage for Sec.~\ref{Sec:s-TRUECARS}, where we show how the same goals can be reached by a single stochastic pulse.

\subsection{The coherent TRUECARS signal}

The TRUECARS technique involves an off-resonant stimulated x-ray Raman process. The pulse $\mathcal{E}_1$ is responsible for the excitation of the system, while $\mathcal{E}_2$ stimulates the emission of the photon. $\mathcal{E}_1$ and $\mathcal{E}_2$ may represent two distinct pulses, or two components of a single broadband pulse. The associated loop diagram \cite{mukamel2010ultrafast} is shown in Fig.~\ref{fig:Feynman}.

The off-resonant stimulated Raman process is described in the rotating-wave approximation by the following effective light--matter interaction Hamiltonian \cite{dorfman2013time},
\begin{equation}
\hat{H}_{\mathrm{int}} = -\hat{\alpha} (\hat{\mathcal{E}}_2^{\dagger}\hat{\mathcal{E}}_1 + \hat{\mathcal{E}}_1^{\dagger}\hat{\mathcal{E}}_2).
\label{eq:intHam}
\end{equation}
Here, $\hat{\alpha}$ is the electronic polarizability operator, while 
\begin{equation}
\hat{\mathcal{E}}_i = \sum_{j_i} \uimm\,\sqrt{\frac{2\pi\omega_{j_i}}{V}}\,\hat{a}(\omega_{j_i})
\label{eq:field}
\end{equation}
and $\hat{\mathcal{E}}_{i}^{\dagger}$, $i\in\{1,\,2\}$, are the positive- and negative-frequency components of the electric-field operator, respectively. In Eq.~(\ref{eq:field}), the index $j_i$ runs over the modes of the $i$th pulse, $V$ is the quantization volume, and $\hat{a}^{\dagger}(\omega)$ and $\hat{a}(\omega)$ are the creation and annihilation operators of a photon with frequency $\omega$, respectively. The effective light--matter interaction Hamiltonian in Eq.~(\ref{eq:intHam}) and the polarizability operator appearing therein can be expressed in terms of molecular charge- and current-density operators. This is shown in Appendix~\ref{App:minimalcoupling}, where we derive the TRUECARS signal starting from the minimal-coupling Hamiltonian of Eq.~(\ref{eq:intHam-minimalcoupling}), which fully describes the radiation--molecule interaction without expansion in multipoles.

The signal $S(\omega_{\mathrm{s}})$ is defined in terms of the frequency-resolved probe-pulse intensity after passing through the sample. It is given by the time-integrated rate of change of the number of photons of frequency $\omega_{\mathrm{s}}$ in the $\mathcal{E}_2$ pulse \cite{marx2008nonlinear},
\begin{equation}
S(\omega_{\mathrm{s}})\diff\omega_{\mathrm{s}} = -\int\diff t\,\left\langle \frac{\diff \hat{N}_2(\omega_{\mathrm{s}})}{\diff t}\right\rangle,
\label{eq:signaldefinition}
\end{equation}
where $\hat{N}_2(\omega_{\mathrm{s}}) = \hat{a}^{\dagger}(\omega_{\mathrm{s}})\hat{a}(\omega_{\mathrm{s}})$ is the number operator of a photon with the detected signal frequency $\omega_{\mathrm{s}}$. This is a heterodyne-detected signal, and the subscript 2 implies that only modes of pulse $\mathcal{E}_2$ are measured. For an optically thin medium, $N_{\mathrm{mol}}\,S(\omega_{\mathrm{s}})\diff \omega_{\mathrm{s}}$ represents the change in the number of x-ray photons detected in the differential frequency interval $[\omega_{\mathrm{s}},\,\omega_{\mathrm{s}} + \diff\omega_{\mathrm{s}}]$ owing to the interaction of $\mathcal{E}_2$ with $N_{\mathrm{mol}}$ molecules. A similar approach, based on the time-integrated rate of energy exchange between light and matter, was used to model attosecond transient-absorption spectroscopy in the presence of strong near-infrared fields \cite{wu2016theory}.

The signal is calculated by starting with the Heisenberg equations of motion for the photon number operator, via the commutator $[\hat{H}_{\mathrm{int}},\,\hat{N}_2(\omega_{\mathrm{s}})]$ and the interaction Hamiltonian in Eq.~(\ref{eq:intHam}). The expectation values over the fields' degrees of freedom are calculated assuming coherent states, thus replacing the field operators $\hat{\mathcal{E}}_i$ with classical electric fields
\begin{equation}
\mathcal{E}_i(t) = E_i(t - T)\,\eu^{-\uimm\omega_{\mathrm{X}i}(t-T)}.
\label{eq:classicalfield}
\end{equation}
$E_i(t)$ are complex envelope functions, the pulses have carrier frequencies $\omega_{\mathrm{X}i}$, and are both centered at the time delay $T$. We further introduce the frequency-domain envelope $\tilde{E}_i(\omega) = \int \diff t\, E_i(t)\,\eu^{\uimm\omega t}$, with $E_i(t) = \int \diff \omega\, \tilde{E}_i(\omega)\,\eu^{-\uimm\omega t}/(2\pi)$. Measuring the frequency-dispersed spectrum $S(\omega_{\mathrm{s}})$ for different time delays $T$ results in the frequency- and time-resolved signal
\begin{equation}
\begin{aligned}
&S(\omega_{\mathrm{s}}, T)\\
=\,& 2\Imm\biggl\{\tilde{E}^*_2(\omega_{\mathrm{s}} - \omega_{\mathrm{X}2})  \int \diff t\,E_1(t-T)\,\eu^{\uimm(\omega_{\mathrm{s}} - \omega_{\mathrm{X}1})(t-T)}\,\langle\hat{\alpha}(t)\rangle\biggr\} \\
=\,&2\Imm\biggl\{\tilde{E}_2^*(\omega_{\mathrm{s}} - \omega_{\mathrm{X}2}) \int\frac{ \diff\omega}{2\pi}\,\tilde{E}_1(\omega_{\mathrm{s}}- \omega_{\mathrm{X}1}- \omega)\,\eu^{-\uimm\omega T}\langle\hat{\tilde{\alpha}}(\omega)\rangle \biggr\},
\label{eq:signal}
\end{aligned}
\end{equation}
with the expectation values of the polarizability operator
\begin{equation}
\begin{aligned}
\langle\hat{\alpha}(t)\rangle &= \Tr\{\hat{\alpha}\hat{\rho}(t)\} = \sum_{i,j}\alpha_{ji}\rho_{ij}(t),\\
\langle\hat{\tilde{\alpha}}(\omega)\rangle &= \Tr\{\hat{\alpha}\hat{\tilde{\rho}}(\omega)\} = \sum_{i,j}\alpha_{ji}\tilde{\rho}_{ij}(\omega),
\end{aligned}
\label{eq:alphas}
\end{equation}
given in terms of the valence-space elements of the density matrix of the system $\hat{\rho}(t)$ or its Fourier transform $\hat{\tilde{\rho}}(\omega)$.

The signal in Eq.~(\ref{eq:signal}) depends on the dynamics of the system via $\langle\hat{\alpha}(t)\rangle$. For sufficiently low pulse intensities, at the level of perturbation theory shown in Fig.~\ref{fig:Feynman}, the TRUECARS signal gives direct access to the free, unperturbed evolution of the molecular electronic and nuclear wavepacket. This is the regime we will focus on in Secs.~\ref{Sec:c-TRUECARS-TDF} and \ref{Sec:s-TRUECARS}, to illustrate how the TRUECARS signal, be it implemented with phase-controlled or stochastic pulses, offers a good joint temporal and spectral resolution---a key requirement for the spectroscopy of ultrafast molecular dynamics. 

By coupling the system to the continuum, the x-ray probe pulse can cause photoionization and ensuing population losses. The influence of photoionization on $\langle\hat{\alpha}(t)\rangle$ can reduce the strength of the signal and erode its temporal and spectral resolution. X-ray fluxes should thus be properly optimized, so that these competing decay losses will not compromise the resolution provided by the TRUECARS technique. Changes in $\langle\hat{\alpha}(t)\rangle$ due to x-ray photoionization or additional higher-order strong-field interactions and their influence on the TRUECARS signal are discussed in Appendix~\ref{App:Photoionization}.

The strength of the TRUECARS signal is determined by the intensity of the x-ray probe pulses in Eq.~(\ref{eq:signal}), the density and size of the molecular sample in the x-ray focal volume, and the amplitude of the polarizability matrix elements $\alpha_{ij}$. Both the signal in Eq.~(\ref{eq:signal}) and the pulse spectral intensity are proportional to the second power of the peak field strength. The relevant signal-to-background ratio, defined as the ratio between the number of absorbed and incident probe-pulse photons, is thus independent of the pulse peak intensity. The maximum pulse intensity can thus be reduced to limit x-ray photoionization without compromising the signal-to-background ratio. At the same time, the signal strength and its ratio to the background pulse intensity can be maximized via a suitable choice of the molecule and by optimizing its density in the experiment. Furthermore, the molecular polarizability can be significantly increased by using x-ray pulses near-resonant to the core-excited states in the molecule, as we discuss in Sec.~\ref{Sec:s-TRUECARS-CoIn} (see, e.g., Fig.~\ref{fig:strengths}). For such regime, we predict in Appendix~\ref{App:signal-to-background} a signal-to-background ratio of $\sim1\%$. We recognize that detecting such signal-to-background ratio may be challenging, especially when using stochastic FEL pulses, but should still be within the capabilities of present and future x-ray detectors. We also notice that stimulated resonant x-ray Raman scattering, the building block of TRUECARS in its resonant x-ray implementation, was already successfully demonstrated in atomic neon \cite{weninger2013stimulatedPRL} in the presence of photoionization channels. Very recently, electronic population transfer following impulsive stimulated resonant x-ray Raman scattering was also observed in NO molecules \cite{oneal2020electronic} thanks to the availability of novel attosecond x-ray FEL pulses. 

\subsection{The c-TRUECARS signal for a model system with a time-dependent frequency}
\label{Sec:c-TRUECARS-TDF}

To illustrate the joint temporal and spectral resolution of c-TRUECARS, we will employ a model system consisting of two electronic states with a time-dependent frequency switching between two values \cite{mukamel2011communication}. This can represent, e.g., photoisomerization. 

\begin{figure}
\centering
\includegraphics[width=0.9\linewidth]{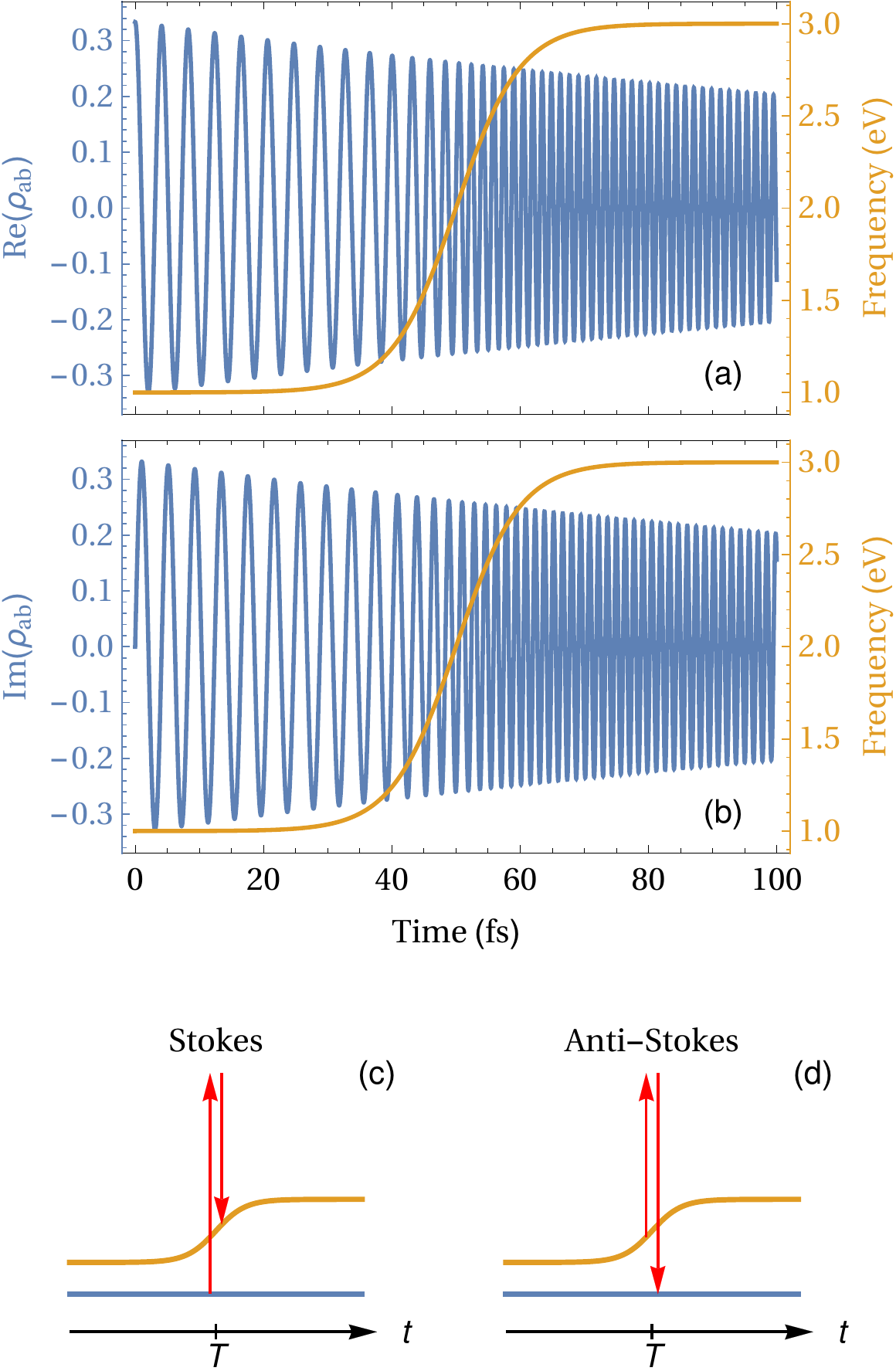}
\caption{Time-dependent-frequency model. (a,b) Evolution of the time-dependent frequency $\omega_{ba}(t)$ [Eq.~(\ref{eq:energies}), yellow curves], along with the (a) real and (b) imaginary parts of the corresponding coherence $\rho_{ab}(t)$ [Eq.~(\ref{eq:coherence}), blue curves]. (c) Stokes- and (d) anti-Stokes-type contributions to the TRUECARS signal between states (blue) $a$ and (yellow) $b$. The red arrows represent the off-resonant stimulated Raman process.}
\label{fig:energiespopcoh}
\end{figure}

We assume a two-level model, with states $a$ and $b$ and a time-dependent frequency
\begin{equation}
\omega_{ba}(t) = \omega_0 + \Delta\omega\,\frac{\tanh[(t-t_0)/\Delta t]}{2},
\label{eq:energies}
\end{equation}
with central frequency $\omega_0 = 2\,\mathrm{eV}$, central time $t_0 = 50\,\mathrm{fs}$, and with a frequency variation of $\Delta\omega = 2\,\mathrm{eV}$ within a time interval of $\Delta t = 10\,\mathrm{fs}$, as shown in Fig.~\ref{fig:energiespopcoh}. The population dynamics are modeled by
\begin{equation}
\rho_{ii}(t) = \rho_{ii,0}\,\eu^{-\gamma_i t},
\end{equation}
with $\rho_{aa,0} = \rho_{bb,0} = 1/3$ and $\gamma_1 = \gamma_2 = 1/(200\,\mathrm{fs})$, and the evolution of the coherences $\rho_{ab}(t)$ and $\rho_{ba}(t) = \rho_{ab}^*(t)$ is given by
\begin{equation}
\rho_{ab}(t) = \sqrt{\rho_{aa}(t)\,\rho_{bb}(t)}\,\eu^{\uimm\int_{-\infty}^{t}\diff t'\,\omega_{ba}(t')}.
\label{eq:coherence}
\end{equation}

\begin{figure*}[t]
\centering
\includegraphics[width=0.95\linewidth]{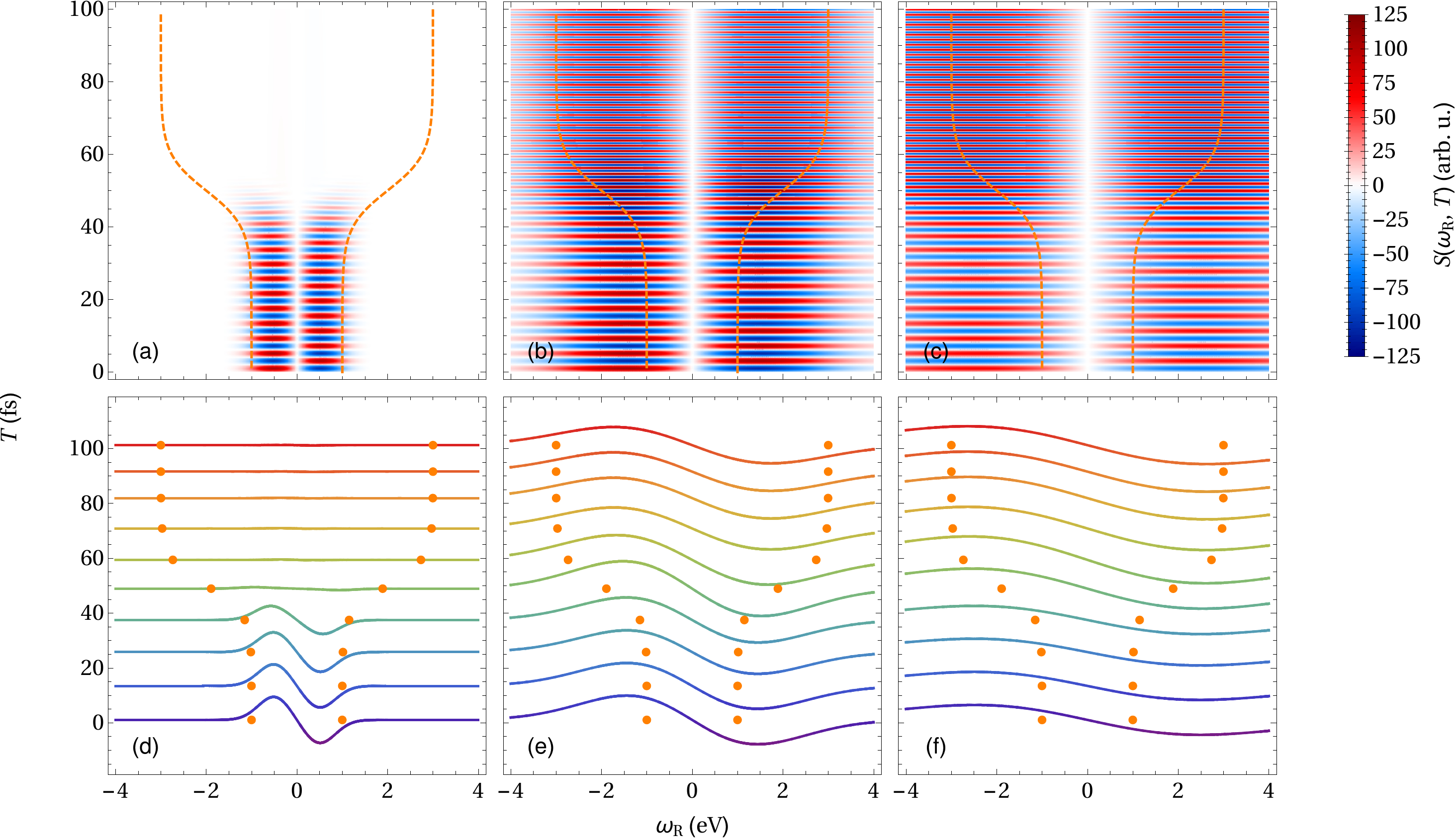}
\caption{c-TRUECARS signal with a single coherent pulse. The signal [Eq.~(\ref{eq:NarrowBroad})] is exhibited for a pulse spectrum $\tilde{E}(\omega)$ of bandwidth (a,d) $\sigma = 0.5\,\mathrm{eV}$ ($1/\sigma = 1.3\,\mathrm{fs}$), (b,e) $\sigma = 2\,\mathrm{eV}$ ($1/\sigma =0.33\,\mathrm{fs}$), and (c,f) $\sigma = 3.5\,\mathrm{eV}$ ($1/\sigma =0.19\,\mathrm{fs}$). The signal is shown (a--c) as a function of the time delay $T$ and the Raman frequency $\omega_{\mathrm{R}}$, and (d--f) for selected time delays. The orange dashed line in (a--c) and the orange dots in (d--f) display the time-dependent frequency $\omega_{ba}(t)$ in Eq.~(\ref{eq:energies}).}
\label{fig:OneCoh}
\end{figure*}

We shall recast the c-TRUECARS signal in terms of the Raman frequency
\begin{equation}
\omega_{\mathrm{R}} = \omega_{\mathrm{s}} - \omega_{\mathrm{X}1}
\end{equation}
and the difference of the x-ray carrier frequencies
\begin{equation}
\omega_{\mathrm{d}} = \omega_{\mathrm{X}1} - \omega_{\mathrm{X}2}
\end{equation}
as
\begin{equation}
\begin{aligned}
&S(\omega_{\mathrm{R}},\omega_{\mathrm{d}}, T)\\
=\,&2\Imm\biggl\{\tilde{E}_2^*(\omega_{\mathrm{R}} + \omega_{\mathrm{d}}) \int\frac{ \diff\omega}{2\pi}\,\tilde{E}_1(\omega_{\mathrm{R}} - \omega)\,\eu^{-\uimm\omega T}\,\langle\hat{\tilde{\alpha}}(\omega)\rangle \biggr\}.
\end{aligned}
\label{eq:NarrowBroad}
\end{equation}
The $\tilde{E}_2(\omega)$ bandwidth determines the spectral detection window, whereas the width of $\tilde{E}_1(\omega)$ sets the time--frequency resolution of the technique. This can be better understood by recasting the signal in the time domain: 
\begin{equation}
\begin{aligned}
&S(\omega_{\mathrm{R}},\omega_{\mathrm{d}}, T)\\
=\,&2\Imm\biggl\{\tilde{E}_2^*(\omega_{\mathrm{R}} + \omega_{\mathrm{d}}) \int \diff t\,E_1(t-T)\,\eu^{\uimm \omega_{\mathrm{R}}(t-T)}\,\langle\hat{\alpha}(t)\rangle \biggr\}.
\end{aligned}
\end{equation}
$E_1(t)$ acts as a temporal gate function centered at time $T$, thereby selecting the dynamics of the system within a time window given by the pulse duration and centered around $T$. The signal is determined by the Fourier transform of this gated function, so that the time duration, i.e., frequency width, of the coherent pulse $E_1(t)$ determines at the same time the temporal and spectral resolutions of the signal.

For a single pulse $E_1(t) = E_2(t) = E(t)$, the signal is quadratic in $E(t)$ and, thus, does not require control over its CEP. However, it does not provide adequate time--frequency resolution. To elucidate why hybrid broad- and narrowband pulses are necessary for the TRUECARS technique of Ref.~\cite{kowalewski2015catching}, Fig.~\ref{fig:OneCoh} shows the c-TRUECARS signal obtained by a single coherent Gaussian pulse
\begin{equation}
\tilde{E}(\omega)=\eu^{-\tfrac{\omega^2}{2\sigma^2}}
\end{equation}
with carrier frequency $\omega_{\mathrm{X}}$ and for different bandwidths $\sigma$. This could be realized experimentally at FEL facilities, e.g., via a split-and-delay module \cite{lu2018development, castagna2013x}. For a narrowband pulse [Fig.~\ref{fig:OneCoh}(a,d)], the observation bandwidth is too narrow to reproduce the change of the frequency from left to right. Increasing the width, from Fig.~\ref{fig:OneCoh}(b,e) to Fig.~\ref{fig:OneCoh}(c,f), offers a broader observation bandwidth, but this is accompanied by a notable decrease in frequency resolution. Shorter pulses provide a narrower time window of the gate function, with improved temporal but significantly deteriorated spectral resolution.

\begin{figure*}[t]
\centering
\includegraphics[width=0.95\linewidth]{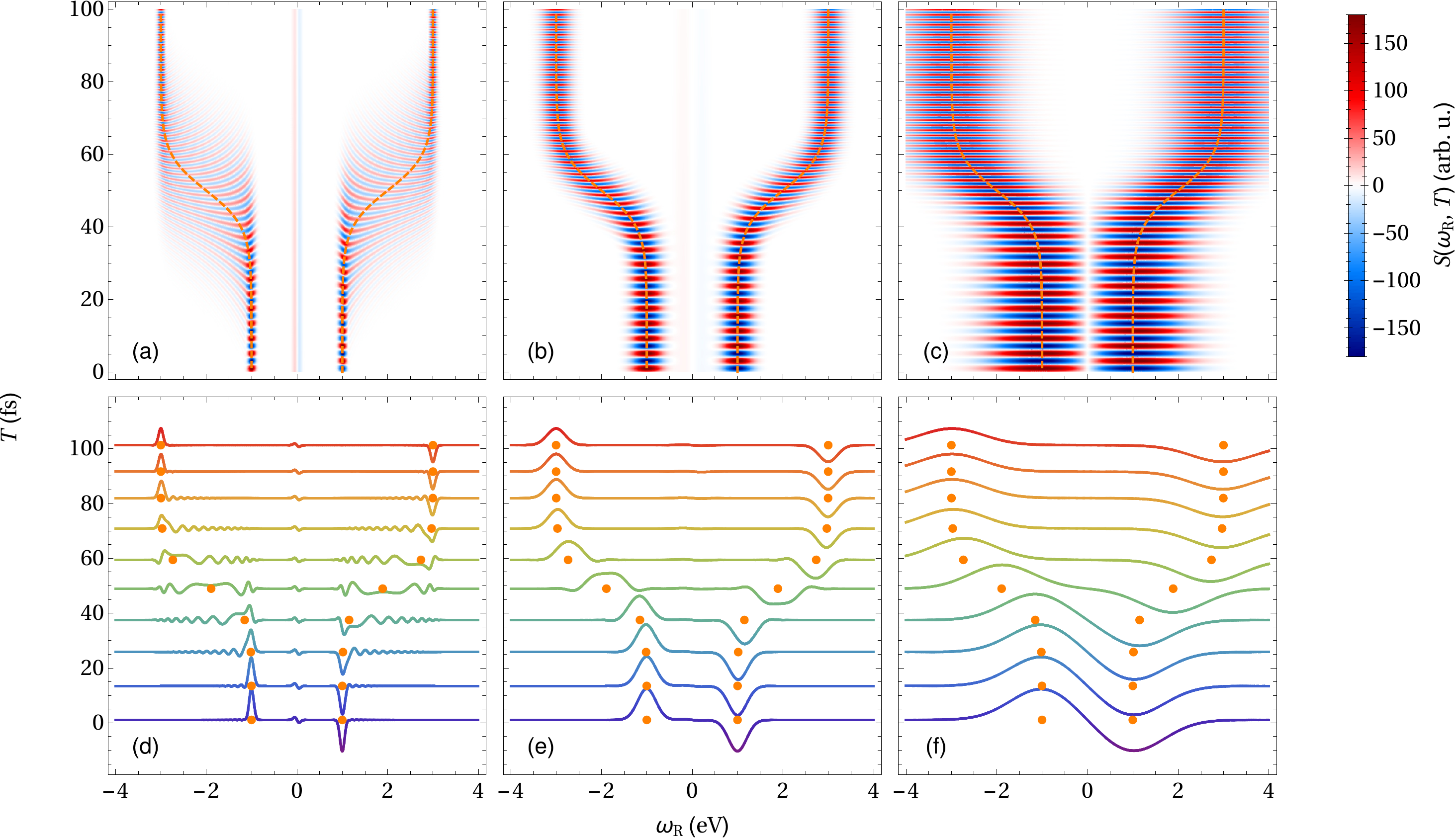}
\caption{c-TRUECARS signal with two coherent phase-controlled pulses. The signal [Eq.~(\ref{eq:NarrowBroad})] is exhibited for a broadband pulse $\tilde{E}_2(\omega)$ of width $\sigma_2 = 10\,\mathrm{eV}$ ($1/\sigma_2 = 66\,\mathrm{as}$) and a narrowband pulse $\tilde{E}_1(\omega)$ of width (a,d) $\sigma_1 = 50\,\mathrm{meV}$ ($1/\sigma_1 = 13\,\mathrm{fs}$), (b,e) $\sigma_1 = 0.2\,\mathrm{eV}$ ($1/\sigma_1 = 3.3\,\mathrm{fs}$), and (c,f) $\sigma_1 = 0.7\,\mathrm{eV}$ ($1/\sigma_1 = 0.94\,\mathrm{fs}$). The signal is shown (a--c) as a function of the time delay $T$ and the Raman frequency $\omega_{\mathrm{R}}$, and (d--f) for selected time delays. The orange dashed line in (a--c) and the orange dots in (d--f) display the time-dependent frequency $\omega_{ba}(t)$ in Eq.~(\ref{eq:energies}). The carrier frequencies of the pulses must be such that $|\omega_{\mathrm{d}}|<\sigma_2$. Here, we set $\omega_{\mathrm{d}} = 0$. Good temporal and spectral resolutions are shown.}
\label{fig:NarrowBroad}
\end{figure*}

Two coherent pulses are thus needed to better control the observation bandwidth and the time--frequency resolution. The arrival time of the pulses is given by their identical time delay $T$. We further set a vanishing CEP difference $(\varphi_2 - \varphi_1) = 0 $ between the two pulses. To ensure that a broad frequency range can be accessed, we use a broadband $\tilde{E}_2(\omega)$. The c-TRUECARS signal thus only weakly depends on $\omega_{\mathrm{d}}$, as long as this lies within the bandwidth $\sigma_2$ of the pulse. By scanning the Raman frequency $\omega_{\mathrm{R}}$, the signal exhibits the appearance of Raman resonances, with a time--frequency resolution determined by the width of $\tilde{E}_1(\omega)$. This is exemplified in Fig.~\ref{fig:NarrowBroad} for two Gaussian pulses
\begin{equation}
\tilde{E}_i(\omega)=\eu^{-\tfrac{\omega^2}{2\sigma_i^2}},
\end{equation}
$i\in\{1,\,2\}$, with a broadband pulse $\tilde{E}_2(\omega)$ and a narrowband pulse $\tilde{E}_1(\omega)$.

The c-TRUECARS signal in Figs.~\ref{fig:NarrowBroad}(a--c) shows contributions at positive and negative Raman frequencies, centered at $+/-$ the local frequency $\omega_{ba}(T)$ of the Raman resonance. Stokes- and anti-Stokes-type processes, shown in Fig.~\ref{fig:energiespopcoh}(c) and \ref{fig:energiespopcoh}(d), respectively, simultaneously contribute to both branches of the signal, leading to absorption or emission of a photon depending on the phase of $\langle\hat{\alpha}(T)\rangle$. A Stokes-type process leads to absorption 
at $\omega_{\mathrm{R}}>0$ and emission 
at $\omega_{\mathrm{R}}<0$, while emission at $\omega_{\mathrm{R}}>0$ and absorption at $\omega_{\mathrm{R}}<0$ are induced by an anti-Stokes process. The oscillatory variation of the signal with $T$, which reveals a time-dependent redistribution of energy from emission to absorption, reflects the molecular polarizability $\langle\hat{\alpha}(T)\rangle$ encountered by the pulses at different time delays. 

For the time-dependent frequency to be imprinted in the signal, the gate function $E_1(t)$ must be sufficiently short compared to the variation time scale of $\omega_{ba}(t)$. However, $E_1(t)$ must be also long enough to include a sufficient number of oscillations of the system at the frequency $\omega_{ba}(t)$ itself, thereby providing frequency resolution. Figures~\ref{fig:NarrowBroad}(a,d) show c-TRUECARS for a long narrowband pulse $E_1(t)$. While the signal provides a good frequency resolution in the regions in which $\omega_{ba}(T)$ is constant, time--frequency resolutions are both lost when the transition frequency is time dependent. The signal results from the average of all molecular Raman frequencies contributing within the long duration of $E_1(t)$, and the local frequency $\omega_{ba}(T)$ cannot be accessed. This can be understood by considering the limiting case of a continuous-wave field, $\tilde{E}_1(\omega) = 2\pi\delta(\omega)$, where the signal reduces to
\begin{equation}
S(\omega_{\mathrm{R}}, \omega_{\mathrm{d}}, T)=2\Imm\left\{\tilde{E}_2^*(\omega_{\mathrm{R}} + \omega_{\mathrm{d}})\,\eu^{-\uimm\omega_{\mathrm{R}} T}\,\langle\hat{\tilde{\alpha}}(\omega_{\mathrm{R}})\rangle  \right\},
\end{equation}
with no temporal information accessed via $T$.

The c-TRUECARS signal is shown in Figs.~\ref{fig:NarrowBroad}(c,f) for a very short pulse $E_1(t)$. As already pointed out while discussing Fig.~\ref{fig:OneCoh}, the very good time resolution achieved in this case is accompanied by a significant erosion of the frequency resolution. The pulse is short compared to the local oscillating period $2\pi/\omega_{ba}(T)$ of the molecule, with a consequent broadening of the Raman peaks in the signal. In the limiting case in which $\tilde{E}_1(\omega) = 1$ (a $\delta$-like excitation in time domain), the signal
\begin{equation}
S(\omega_{\mathrm{R}}, \omega_{\mathrm{d}}, T)=2 \Imm\left\{\tilde{E}_2^*(\omega_{\mathrm{R}} + \omega_{\mathrm{d}})\,\langle\hat{\alpha}(T)\rangle \right\}
\end{equation}
can monitor the time evolution of the system, but with no frequency information. We notice that $\langle\hat{\alpha}(T)\rangle$ is in general a real quantity, and the signal will vanish for such very broadband pulses, if they both have the same CEP. 

A trade-off between the two molecular time scales is thus necessary in order to extract the evolution of $\omega_{ba}(t)$ from the c-TRUECARS signal with optimal joint time--frequency resolution. This case is depicted in Figs.~\ref{fig:NarrowBroad}(b,e).

The populations $\rho_{ii}(t)$ do not carry a dynamical phase. For any bandwidth of $\tilde{E}_1(\omega)$, they do not give rise to a real term in Eq.~(\ref{eq:NarrowBroad}) and do not contribute to the signal. c-TRUECARS can thus directly access the emergence and evolution of the molecular coherences $\rho_{ij}(t)$ in a background-free manner---a crucial requirement for the unambiguous observation of CoIns, as will be shown in Sec.~\ref{Sec:s-TRUECARS-CoIn}.

\section{TRUECARS with a stochastic x-ray pulse}
\label{Sec:s-TRUECARS}

The duration and bandwidth of a coherent pulse are linked by Fourier uncertainty. Therefore, a large observation bandwidth and a controllable time--frequency resolution cannot be achieved by a single coherent pulse, as shown in Fig.~\ref{fig:OneCoh}. Two pulses with a controlled CEP are needed for the implementation of c-TRUECARS. Variations of the pulse CEPs will lead to changes in the signal, which average out to zero. This has hindered the implementation of the c-TRUECARS technique with currently available stochastic FEL pulses. 

When implemented with a single SASE FEL pulse, the TRUECARS signal itself [Eq.~(\ref{eq:signal})] is a stochastic process. However, since each signal is uniquely related to the particular pulse producing it, valuable spectroscopic information can be retrieved by exploiting the correlations between the spectral components of the stochastic pulse \cite{kimberg2016stochastic, tollerud2019femtosecond, asban2019frequency, osipov2019time, kayser2019core}. In the following, we will investigate the s-TRUECARS signal defined by the correlation between the stimulated Raman signal $S(\omega_{\mathrm{s}2},T)$, induced by a given stochastic pulse $E(t)$, and the spectral intensity $|\tilde{E}(\omega_{\mathrm{s}1} - \omega_{\mathrm{X}})|^2$ of that same pulse. The s-TRUECARS signal, obtained by averaging this correlation function over independent realizations of the stochastic process, provides time--frequency resolution over a broad bandwidth, thus enabling the observation of fast molecular dynamics, such as at CoIns, with current x-ray FEL pulses. The s-TRUECARS signal shares the advantages of c-TRUECARS, as it enables background-free access to the evolution of the coherences in the system. 
However, it does not require any control over the pulse spectral phase.

\subsection{The s-TRUECARS correlation function}

The s-TRUECARS technique exploits the stimulated off-resonant Raman scattering of a stochastic pulse off the system. The signal is given by Eq.~(\ref{eq:signal}) where $E_1(t) = E_2(t) = E(t)$ is the envelope of the stochastic pulse and $\omega_{\mathrm{X}1} = \omega_{\mathrm{X}2} = \omega_{\mathrm{X}}$ its carrier frequency. Information with time and frequency resolution is extracted by correlating each signal with the pulse producing it, and then averaging over independent realizations of the stochastic pulse. 

We thus introduce the covariance signal \cite{kimberg2016stochastic, tollerud2019femtosecond, asban2019frequency, osipov2019time} given by the correlation function between the pulse spectral intensity at frequency $\omega_{\mathrm{s}1}$ and the signal at a different frequency $\omega_{\mathrm{s}2}$:
\begin{equation}
\begin{aligned}
C(\omega_{\mathrm{s}1}, \omega_{\mathrm{s}2}, T) = \,&\langle|\tilde{E}(\omega_{\mathrm{s}1} - \omega_{\mathrm{X}})|^2 \,S(\omega_{\mathrm{s}2}, T)\rangle \\
&- \langle|\tilde{E}(\omega_{\mathrm{s}1} - \omega_{\mathrm{X}})|^2\rangle\langle S(\omega_{\mathrm{s}2}, T)\rangle.
\end{aligned}
\label{eq:correlationearlydefinition}
\end{equation}
Here, $\langle\cdots\rangle$ denotes the average over independent measurements. By using Eq.~(\ref{eq:signal}), the correlation function expressed in terms of the frequency differences 
\begin{equation}
\omega'_{\mathrm{s}i} = \omega_{\mathrm{s}i} - \omega_{\mathrm{X}}
\end{equation}
reduces to
\begin{equation}
C(\omega'_{\mathrm{s}1}, \omega'_{\mathrm{s}2}, T) = 2\Imm\biggl\{\int\frac{ \diff\omega}{2\pi}\,  G(\omega'_{\mathrm{s}1}, \omega'_{\mathrm{s}2}, \omega)\,\eu^{-\uimm\omega T}\,\langle\hat{\tilde{\alpha}}(\omega)\rangle \biggr\},
\label{eq:correlationdefinition}
\end{equation}
with 
\begin{equation}
\begin{aligned}
G(\omega'_{\mathrm{s}1}, \omega'_{\mathrm{s}2}, \omega) =\,& F_4(\omega'_{\mathrm{s}1},\omega'_{\mathrm{s}1}, \omega'_{\mathrm{s}2}, \omega'_{\mathrm{s}2} - \omega) \\
&- F_2(\omega'_{\mathrm{s}1},\omega'_{\mathrm{s}1})\,F_2(\omega'_{\mathrm{s}2}, \omega'_{\mathrm{s}2} - \omega)
\end{aligned}
\label{eq:stochgatefreq}
\end{equation}
defined in terms of the two- and four-point correlation functions of the field $\tilde{E}(\omega)$. The signals and the pulse spectral intensities are correlated at frequencies $\omega_{\mathrm{s}2}$ and $\omega_{\mathrm{s}1}$, with the frequency difference $(\omega_{\mathrm{s}2} - \omega_{\mathrm{s}1})$ here playing the role of the Raman frequency---in c-TRUECARS it was $(\omega_{\mathrm{s}} - \omega_{\mathrm{X}1})$. The frequency $\omega_{\mathrm{s}1}$ thus provides the reference necessary to reveal the Raman resonances in the molecule and the evolution of their time-dependent frequencies by scanning $\omega_{\mathrm{s}2}$.

\subsection{The s-TRUECARS signal for the time-dependent-frequency model}
\label{Sec:s-TRUECARS-TDF}

We first consider a stochastic FEL pulse based on the UDSP model with $a = \pi$. We calculate $G(\omega'_{\mathrm{s}1}, \omega'_{\mathrm{s}2}, \omega)$ based on Eqs.~(\ref{eq:F2uniformstat}) and (\ref{eq:F4uniformstat}) with $s(\pi) = 0$, so that the correlation function in Eq.~(\ref{eq:correlationdefinition}) reduces to
\begin{widetext}
\begin{equation}
\begin{aligned}
C(\omega'_{\mathrm{s}1},\omega'_{\mathrm{s}2}, T) 
=\,& 2\varLambda^2\pi\tau^2\,\Imm\biggl\{ |\tilde{g}(\omega'_{\mathrm{s}1})|^2\,\tilde{g}^*(\omega'_{\mathrm{s}2})\,\eu^{-\tfrac{(\omega'_{\mathrm{s}2}-\omega'_{\mathrm{s}1})^2\tau^2}{4}} \int\frac{ \diff\omega}{2\pi} \,\tilde{g}(\omega'_{\mathrm{s}2}-\omega)\,\eu^{-\tfrac{(\omega'_{\mathrm{s}2}-\omega'_{\mathrm{s}1} - \omega)^2\tau^2}{4}}\,\eu^{-\uimm\omega T}\,\langle\hat{\tilde{\alpha}}(\omega)\rangle \biggr\}\\
=\,&2\varLambda^2\pi\tau^2\,\Imm\biggl\{ \Bigl|\tilde{g}\Bigl(\omega_{\mathrm{m}}-\frac{\omega_{\mathrm{R}}}{2}\Bigr)\Bigr|^2\,\tilde{g}^*\Bigl(\omega_{\mathrm{m}}+\frac{\omega_{\mathrm{R}}}{2}\Bigr)\,\eu^{-\tfrac{\omega_{\mathrm{R}}^2\tau^2}{4}} \int\frac{ \diff\omega}{2\pi}\,\tilde{g}\Bigl(\omega_{\mathrm{m}}+\frac{\omega_{\mathrm{R}}}{2}-\omega\Bigr)\,\eu^{-\tfrac{(\omega_{\mathrm{R}}- \omega)^2\tau^2}{4}}\,\eu^{-\uimm\omega T}\,\langle\hat{\tilde{\alpha}}(\omega)\rangle \biggr\},
\end{aligned}
\end{equation}
\end{widetext}
where we have introduced the Raman frequency
\begin{equation}
\omega_{\mathrm{R}} = \omega'_{\mathrm{s}2} - \omega'_{\mathrm{s}1} = \omega_{\mathrm{s}2}-\omega_{\mathrm{s}1}
\end{equation}
and the mean detected signal frequency
\begin{equation}
\omega_{\mathrm{m}} = \frac{\omega'_{\mathrm{s}2} + \omega'_{\mathrm{s}1}}{2} = \frac{\omega_{\mathrm{s}2} + \omega_{\mathrm{s}1}}{2} -\omega_{\mathrm{X}}.
\end{equation}
The correlation function displays the same structure as the c-TRUECARS signal implemented with a single pulse. This can be more clearly seen by considering the limit of an extremely broadband frequency envelope, $\tilde{g}(\omega)\rightarrow 1$, for which the correlation function reads
\begin{equation}
\begin{aligned}
&C(\omega_{\mathrm{R}},\omega_{\mathrm{m}}, T) \\
=\,& 2\varLambda^2\pi\tau^2\,\Imm\biggl\{\eu^{-\tfrac{\omega_{\mathrm{R}}^2\tau^2}{4}} \int\frac{ \diff\omega}{2\pi} \,\eu^{-\tfrac{(\omega_{\mathrm{R}} - \omega)^2\tau^2}{4}}\,\eu^{-\uimm\omega T}\,\langle\hat{\tilde{\alpha}}(\omega)\rangle \biggr\}.
\end{aligned}
\label{eq:correlationfunctionMaxVariance}
\end{equation}
As apparent in Eq.~(\ref{eq:correlationfunctionMaxVariance}), in spite of the broadband frequency envelope $\tilde{g}(\omega)$, the observation bandwidth of the correlation function is given by the Fourier transform $\tilde{u}(\omega_{\mathrm{R}}/\sqrt{2}) = \tau\,\eu^{-\omega_{\mathrm{R}}^2\tau^2/4}$ of the time envelope $u(t)$. Since the same function also determines the time--frequency resolution of the technique, this leads to the same limitations shown in Fig.~\ref{fig:OneCoh}.

The stochastic UDSP pulse considered above, with a relatively long time envelope $u(t)$ and with $a = \pi$, does not possess a well defined central time $T$. This is apparent in Fig.~\ref{fig:PulseProfileMaxVariance}(a): the pulse features a series of peaks randomly distributed within its duration $\tau$, resulting in a large uncertainty over the position of its central time. However, for the UDSP model with $a<\pi$ and for the composite stochastic pulse of Eq.~(\ref{eq:sumpulse}), a central peak emerges in the pulse temporal envelope [see, e.g., Fig.~\ref{fig:PulseProfileSmallVariance}(a)]. This is crucial to simultaneously utilize the large bandwidth and the long duration of the stochastic pulse, and thus for the implementation of s-TRUECARS, as shown in the following. 

For stochastic UDSP pulses with $a<\pi$ and, thus, $s(a)\neq 0$, the function $G(\omega'_{\mathrm{s}1}, \omega'_{\mathrm{s}2}, \omega)$ can be calculated via Eqs.~(\ref{eq:F2uniformstat}) and (\ref{eq:F4uniformstat}) to first (leading) order in $(\varLambda\tau)$, and the associated correlation function $C(\omega_{\mathrm{R}}, \omega'_{\mathrm{m}}, T) $, in terms of the above introduced Raman and mean frequencies, can be recast in the form
\begin{widetext}
\begin{equation}
\begin{aligned}
C(\omega_{\mathrm{R}} ,\omega_{\mathrm{m}} , T) \,&= 4\pi s^2(a)\left[1+s(a)\,c(a)-2s^2(a)\right]\varLambda\Imm\biggl\{ \Bigl|\tilde{g}\Bigl(\omega_{\mathrm{m}}-\frac{\omega_{\mathrm{R}}}{2}\Bigr)\Bigr|^2\,\tilde{g}^*\Bigl(\omega_{\mathrm{m}}+\frac{\omega_{\mathrm{R}}}{2}\Bigr)\\
&\ \ \ \times \int\frac{ \diff\omega}{2\pi} \,\tilde{g}\Bigl(\omega_{\mathrm{m}}+\frac{\omega_{\mathrm{R}}}{2}-\omega\Bigr)\,\sqrt{\pi}\tau\,\biggl(\eu^{-\tfrac{\omega_{\mathrm{R}2}^2\tau^2}{4}} + \eu^{-\tfrac{(\omega_{\mathrm{R}}- \omega)^2\tau^2}{4}}\biggr)\,\eu^{-\uimm\omega T}\,\langle\hat{\tilde{\alpha}}(\omega)\rangle \biggr\}, 
\end{aligned}
\label{eq:correlationfunction}
\end{equation}
where $s(a) = \sinc(a) = \sin(a)/a$ and $c(a) = \cos(a)$. The properties of the correlation function and the origin of the time--frequency resolution provided by s-TRUECARS can be better understood by writing the integral in Eq.~(\ref{eq:correlationfunction}) in time domain,
\begin{equation}
\begin{aligned}
C(\omega_{\mathrm{R}} ,\omega_{\mathrm{m}} , T) \,&=4\pi s^2(a)\left[1+s(a)\,c(a)-2s^2(a)\right]\varLambda\, \Imm\biggl\{ \Bigl|\tilde{g}\Bigl(\omega_{\mathrm{m}}-\frac{\omega_{\mathrm{R}}}{2}\Bigr)\Bigr|^2\,\tilde{g}^*\Bigl(\omega_{\mathrm{m}}+\frac{\omega_{\mathrm{R}}}{2}\Bigr)\\
&\ \ \ \times \int \diff t \,\biggl[g(t-T)\,\eu^{\uimm\left(\omega_{\mathrm{m}}+\frac{\omega_{\mathrm{R}}}{2}\right)(t-T)}\,\eu^{-\tfrac{\omega_{\mathrm{R}}^2\tau^2}{4}} + \int\diff t'\,g(t')\,\eu^{\uimm\left(\omega_{\mathrm{m}}+\frac{\omega_{\mathrm{R}}}{2}\right)t'}\,\eu^{-\tfrac{(t-t'-T)^2}{\tau^2}}\,\eu^{\uimm\omega_{\mathrm{R}}(t - t' - T)}\biggr]\,\langle\hat{\alpha}(t)\rangle \biggr\},
\end{aligned}
\end{equation}
\end{widetext}
where $g(t)$ is the Fourier transform of the broadband frequency envelope.

\begin{figure*}[t]
\centering
\includegraphics[width=0.95\linewidth]{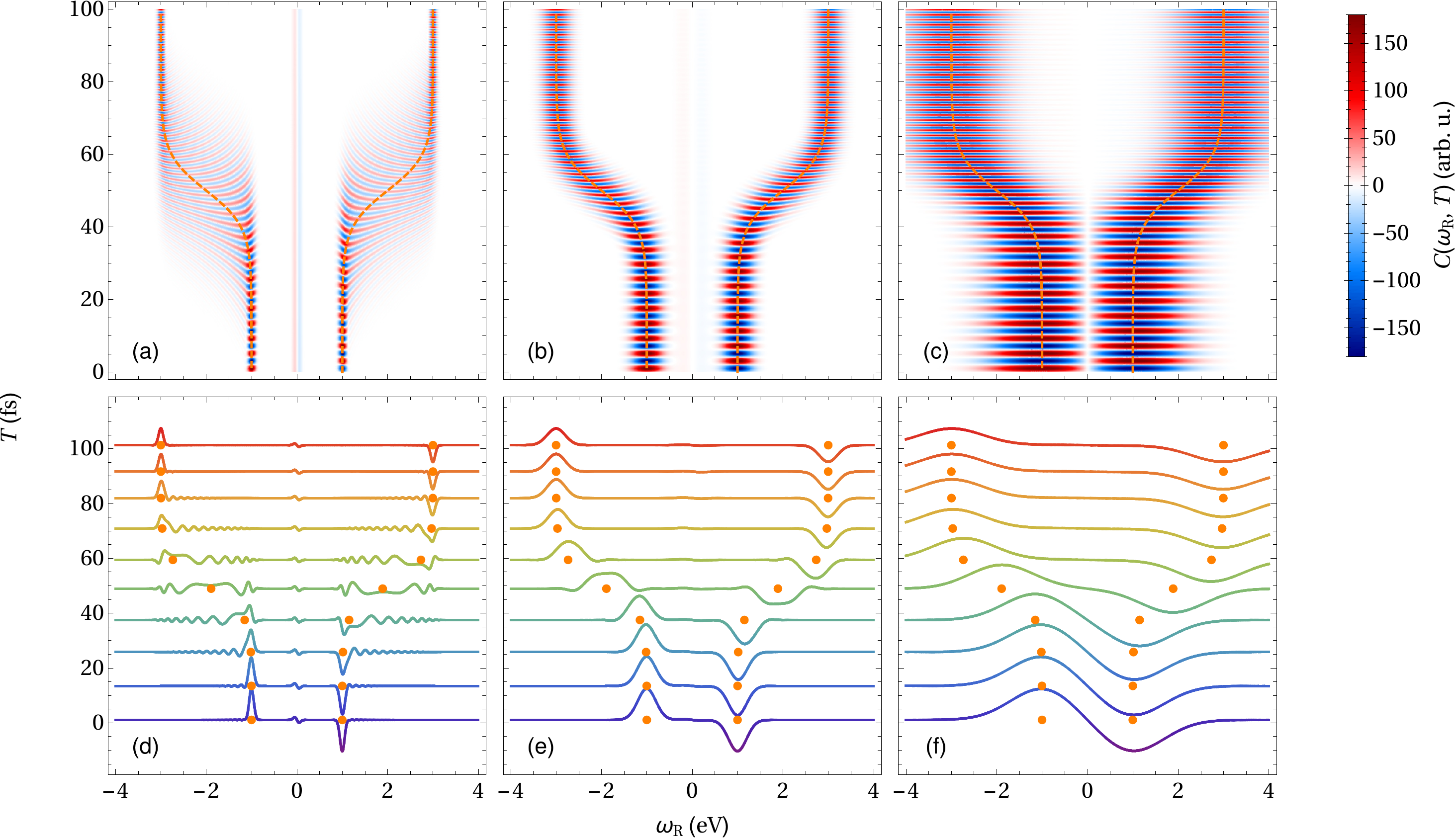}
\caption{s-TRUECARS signal for stochastic x-ray UDSP pulses with $a<\pi$. The correlation function in Eq.~(\ref{eq:correlationfunction}) is exhibited for $\sigma = 10\,\mathrm{eV}$ and (a,d) $\tau = 19\,\mathrm{fs}$, (b,e) $\tau = 4.7\,\mathrm{fs}$, and (c,f) $\tau = 1.3\,\mathrm{fs}$. The correlation function is shown (a--c) as a function of the time delay $T$ and the Raman frequency $\omega_{\mathrm{R}}$, and (d--f) for selected time delays. The orange dashed line in (a--c) and the orange dots in (d--f) display the time-dependent frequency $\omega_{ba}(t)$ in Eq.~(\ref{eq:energies}). The mean frequency $\omega_{\mathrm{m}}$ must be such that $|\omega_{\mathrm{m}}|<\sigma$. Here, we set $\omega_{\mathrm{m}} = 0$.}
\label{fig:StochXFEL}
\end{figure*}

For broadband pulses, the correlation function is virtually independent of the mean frequency $\omega_{\mathrm{m}}$ as long as this lies within the large pulse bandwidth $\sigma$. By scanning the Raman frequency $\omega_{\mathrm{R}}$, the correlation function reveals the appearance of Raman resonances in the system. This is shown in Fig.~\ref{fig:StochXFEL} for a model with time-dependent frequencies and for different pulse durations $\tau$. s-TRUECARS provides the same combination of large observation bandwidth and optimal joint time--frequency resolution enabled by c-TRUECARS, without requiring any phase control of the pulse.

Clear analogies can be drawn between the coherent and stochastic techniques. In the s-TRUECARS correlation function of Eq.~(\ref{eq:correlationfunction}), the frequency envelope $\tilde{g}(\omega)$ sets the observation bandwidth. This is analogous to the role played by $\tilde{E}_2(\omega)$ for c-TRUECARS. The overall frequency envelope of the stochastic pulse should therefore be broad to ensure a wide observation range. The time--frequency resolution of the technique is then determined by the integrand in Eq.~(\ref{eq:correlationfunction}). To better understand this property, it is useful to focus on the limiting case of an extremely broadband pulse, $\tilde{g}(\omega)\rightarrow 1$, where the correlation function only depends on the Raman frequency $\omega_{\mathrm{R}}$ and reduces to
\begin{equation}
\begin{aligned}
&C(\omega_{\mathrm{R}} ,\omega_{\mathrm{m}} , T)\xrightarrow[\tilde{g}(\omega)\rightarrow 1]{}
C(\omega_{\mathrm{R}} , T)\\
\propto\,&\sqrt{\pi}\tau\,\eu^{-\tfrac{\omega_{\mathrm{R}}^2\tau^2}{4}}\,\Imm\biggl\{\int\frac{ \diff\omega}{2\pi}\,\,\eu^{-\uimm\omega T}\,\langle\hat{\tilde{\alpha}}(\omega)\rangle \biggr\}\\
&+\Imm\biggl\{\int\frac{ \diff\omega}{2\pi}\,\sqrt{\pi}\tau\,\eu^{-\tfrac{(\omega_{\mathrm{R}}- \omega)^2\tau^2}{4}}\,\eu^{-\uimm\omega T}\,\langle\hat{\tilde{\alpha}}(\omega)\rangle \biggr\}.
\end{aligned}
\label{eq:stc1}
\end{equation}
The first addend in Eq.~(\ref{eq:stc1}) is proportional to $\Imm\left\{\langle\hat{\alpha}(T)\rangle \right\}$ and vanishes exactly in the limit of a very broadband envelope $\tilde{g}(\omega)$ since $\langle\hat{\alpha}(T)\rangle$ is real. The main contribution to the correlation function thus comes from the second term given by
\begin{equation}
\begin{aligned}
&\Imm\biggl\{ \int\frac{ \diff\omega}{2\pi}\,\sqrt{\pi}\tau\,\eu^{-\tfrac{(\omega_{\mathrm{R}}- \omega)^2\tau^2}{4}}\,\eu^{-\uimm\omega T}\,\langle\hat{\tilde{\alpha}}(\omega)\rangle \biggr\}\\
=\,&\Imm\biggl\{ \int \diff t \,\eu^{-\tfrac{(t-T)^2}{\tau^2}}\,\eu^{\uimm\omega_{\mathrm{R}}(t-T)}\,\langle\hat{\alpha}(t)\rangle \biggr\}.
\end{aligned}
\label{eq:stc2}
\end{equation}
The overall time envelope $|u(t)|^2 = \eu^{-t^2/\tau^2}/(2\pi)$ of the stochastic pulse acts as a gate centered at $t=T$, selecting the dynamics of the system only within a time window $\tau$ centered around $T$. The correlation function results from the Fourier transform of this gated function, with $\tau$ controlling the time--frequency resolution. This is illustrated in Fig.~\ref{fig:StochXFEL} for different stochastic-pulse durations. The role played by the time envelope $|u(t)|^2$ in s-TRUECARS is thus completely analogous to the role of $E_1(t)$ in c-TRUECARS.

The s-TRUECARS signal in Eq.~(\ref{eq:correlationfunction}) was calculated for the stochastic pulses of Eqs.~(\ref{eq:stochpulset}) and (\ref{eq:stochpulse}), based on the UDSP model with $a<\pi$. Such pulses provide a broadband frequency envelope $\tilde{g}(\omega)$, a long time envelope $u(t)$, and a precisely defined central time $T$. All these features are required to achieve large observation widths and a controllable time--frequency resolution. The stochastic pulse presented in Eq.~(\ref{eq:sumpulse}), consisting of a short peaked pulse and a long broadband stochastic UDSP FEL pulse with $a = \pi$, provides the same favourable combination of parameters. As shown in Appendix~\ref{Appendix:C+S}, the corresponding correlation function, given in Eq.~(\ref{eq:correlationfunctionsumpulse}), exhibits exactly the same structure as Eq.~(\ref{eq:correlationfunction}). This composite stochastic x-ray pulse can thus identically enable large observation widths and time--frequency resolutions, without requiring any shaping or control of the pulse phase.

\subsection{s-TRUECARS signal of a conical intersection in the RNA base Uracil}
\label{Sec:s-TRUECARS-CoIn}

The passage through a CoIn of electronic states is a particularly intriguing example of nonadiabatic molecular dynamics originating from the strong coupling of electronic and nuclear degrees of freedom \cite{worth2004beyond,domcke2011conical}. CoIns are electronic degenerate regions of two potential energy surfaces, where electronic and nuclear frequencies become comparable and the Born--Oppenheimer approximation breaks down \cite{born1927quantentheorie}. In spite of being ubiquitous in molecules, CoIns could not be observed directly in an experiment. This is due to the fact that the passage of a molecular wave packet (WP) through CoIns simultaneously involves ultrafast dynamics and very small frequencies, with challenging requirements on the time and frequency resolutions necessary for their observation.

\begin{figure}[t]
	\noindent\begin{centering}
		\includegraphics[width=\linewidth]{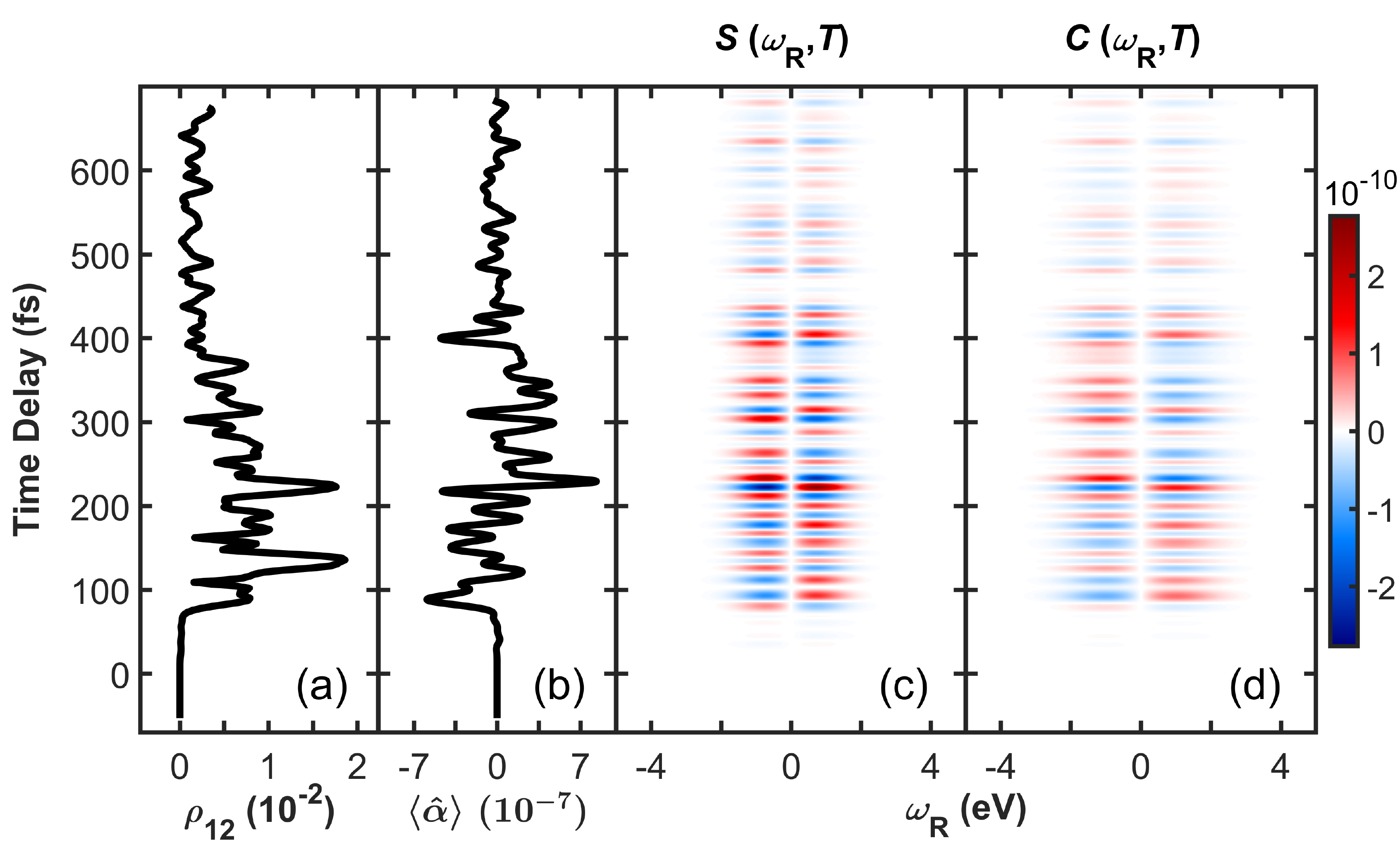}
		\par\end{centering}
	\caption{\stc and \ctc of the Uracil CoIn. The effective Hamiltonian that models the nuclear WP dynamics during relaxation from the \stwo to the \sone state has been described in Ref.~\cite{Keefer2020}. The polarizability [Eq.~(\ref{eq:polarizability})] and the isotropic signal, averaged over the field polarization direction, are exhibited for an off-resonant x-ray pulse of frequency $\omega_{\mathrm{X}}= 354\,\mathrm{eV}$. (a) Magnitude of the coherence between the \stwo and \sone state. After an initial free evolution time in the \stwo state, the nuclear WP reaches the CoIn, and a coherence $\rho_{12}$ emerges due to the bifurcation in the nonadiabatic passage. (b) Expectation value of the polarizability operator resulting from the dynamics in atomic units. (c) \ctc signal according to Eq.~(\ref{eq:NarrowBroad}) using a broadband (attosecond) and a narrowband (femtosecond) x-ray probing field, requiring phase control. (d) \stc signal according to Eqs.~(\ref{eq:stc1})~and~(\ref{eq:stc2}), using a single x-ray probe field with $\tau = 0.93\,\mathrm{fs}$ ($1/\tau = 0.71\,\mathrm{eV}$) and random phase.}
	\label{fig:signal}
\end{figure}

We demonstrate the \stc signal for the photorelaxation of Uracil through a CoIn seam. Uracil is an RNA nucleobase exhibiting ultrafast (femtosecond) relaxation after optical excitation to the bright \stwo state. Due to its biological relevance, interesting photophysics, convenient size and chemical handleability, it is a frequent subject of experimental and theoretical studies, and a promising candidate for pioneering x-ray FEL experiments. An effective Hamiltonian necessary for performing exact nuclear quantum dynamics according to the time-dependent Schr\"odinger equation has been described in Refs.~\cite{Keefer2017,Keefer2020}. It contains two nuclear degrees of freedom and the ground and first two electronically excited states, with a CoIn seam between the \stwo and \sone states. Using a 20~fs full-width at half maximum (FWHM) optical pump in resonance with the bright \szero to \stwo transition, there is a free evolution period of the nuclear WP in the \stwo state. Starting at 100~fs, tails of the WP constantly reach the S\textsubscript{2}/S\textsubscript{1} CoIn region, where it bifurcates and relaxes to the \sone state. For a more detailed description and visualization of this process, see Refs.~\cite{Keefer2017,Keefer2020}. The time-dependent material quantity that is probed by the \tc signal is the vibronic coherence emerging  at the CoIn due to the WP bifurcation. Figure~\ref{fig:signal}(a) displays the magnitude of this coherence. It is initially zero, since only the \stwo state is bright, and becomes finite at around 1000~fs, where the WP reaches the CoIn and the nonadiabatic passage starts. After 250~fs, the coherence magnitude constantly decreases, since major parts of the WP have already evolved away from the CoIn in the \sone state.   

The \ctc signal [Eq.~(\ref{eq:NarrowBroad})] using a broadband (500~as) $\tilde{E}_2^\ast(\omega)$ and a narrowband (2~fs) $E_1(t)$ x-ray pulse to probe this process in Uracil was described in Ref.~\cite{Keefer2020}. The signal was demonstrated to reveal deep insight into the CoIn passage by directly mapping the path of the WP coherence around the CoIn. A major difficulty in potential experimental realizations is that precise phase control of the two probe pulses is required to measure the signal, which is not feasible yet. 

\begin{figure}[t]
	\noindent\begin{centering}
		\includegraphics[width=\linewidth]{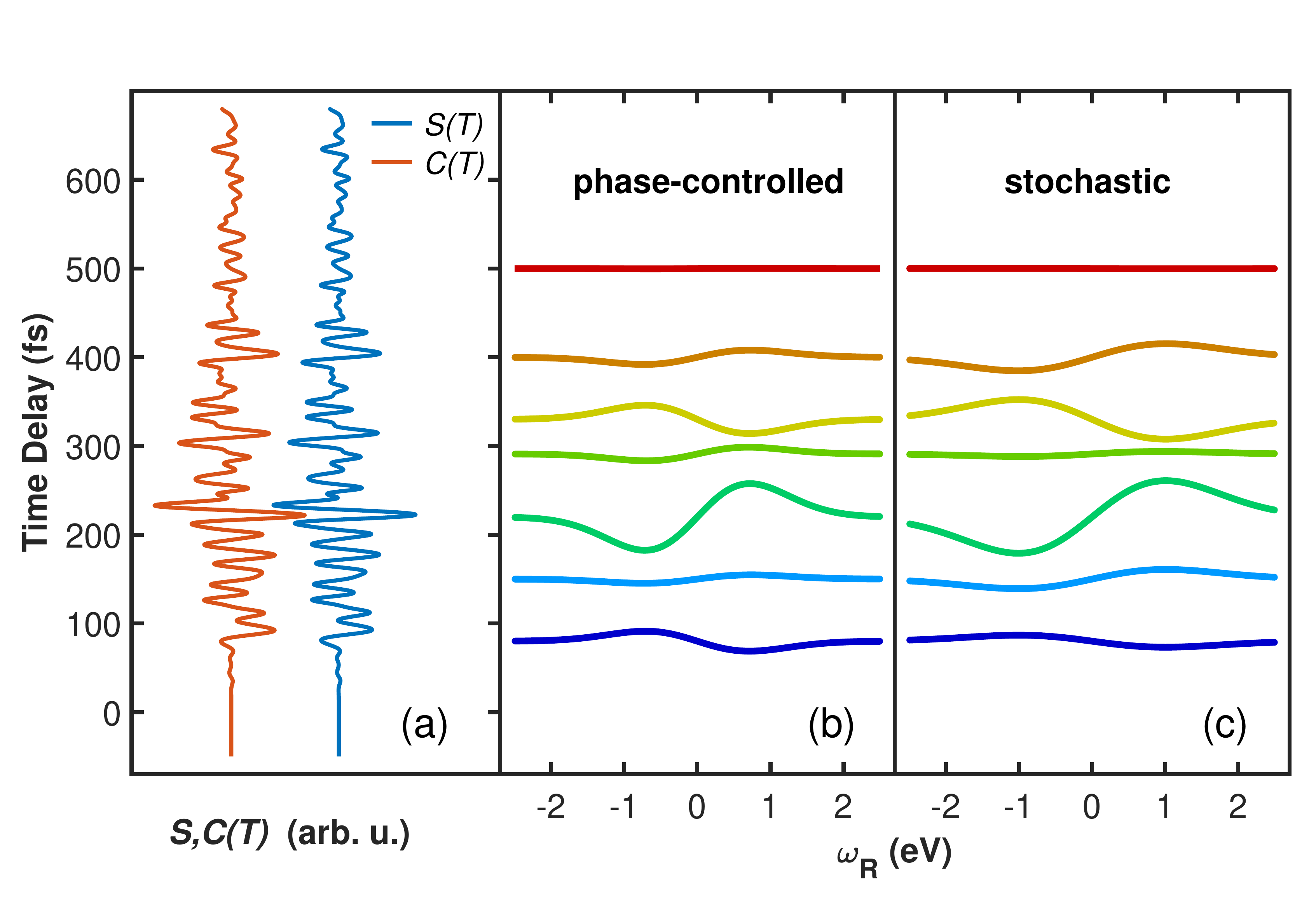}
		\par\end{centering}
	\caption{Horizontal and vertical slices through the signals shown in Fig.~\ref{fig:signal}. (a) Temporal trace at constant Raman frequency $\omega_\mathrm{R}$ corresponding to the maximum signal intensity. The blue line corresponds to the \ctc signal in Fig.~\ref{fig:signal}(c) at $\omega_{\mathrm{R}}$~=~0.7~eV, while the orange line is the \stc signal in Fig.~\ref{fig:signal}(d) at $\omega_{\mathrm{R}}$~=~1.0~eV. The traces are normalized with respect to the maximum signal intensity. (b,c) Frequency slices at constant times for (b) c-TRUECARS and (c) s-TRUECARS. The traces are normalized with respect to the maximum signal intensity and show similar features. Small differences are observed, e.g., at 290~fs, where the \ctc signal exhibits the characteristic gain and loss contributions, while the \stc signal is very close to zero, or at 330~and~400~fs.}
	\label{fig:slices}
\end{figure}

Here, we report the \stc signal according to Eqs.~(\ref{eq:stc1})~and~(\ref{eq:stc2}) using a single stochastic probe pulse rather than two phase-controlled pulses. As demonstrated in Fig.~\ref{fig:signal}, the signal can be measured with almost equivalent detail, and the same physical effects can be resolved. Figure~\ref{fig:signal}(c) displays the signal using the phase-controlled hybrid broad-/narrowband probing scheme. Originally, this pulse configuration was chosen to provide the optimal joint time--frequency resolution which is needed to monitor the ultrafast coherences during the CoIn passage. The \stc signal using a single stochastic x-ray pulse with random phase, as generated from an FEL, is shown in Fig.~\ref{fig:signal}(d). It exhibits the same characteristic oscillations between Stokes 
and anti-Stokes 
contributions. To corroborate this strong similarity, horizontal and vertical cuts through the signal are displayed in Fig.~\ref{fig:slices}. A similar behavior is observed, with some small differences occurring, e.g., in the vertical cut at 290~fs, where the \ctc signal exhibits a small-amplitude oscillation period, while the \stc is very close to zero. Note that Fig.~\ref{fig:signal} shows the isotropic signal, i.e. there is no molecular orientation necessary in the experiment.

Additional physical information about the molecule can be accessed with the \tc signal. The oscillations between blue and red in the frequency-resolved signal are due to the S\textsubscript{2}/S\textsubscript{1} coherence propagating with a dynamical phase owing to the energy difference in the vibronic states \cite{kowalewski2015catching,Keefer2020}. This is also shown in Fig.~\ref{fig:slices}(a) in the horizontal signal slices at constant Raman frequency $\omega_{\mathrm{R}}$. The energy splitting between the contributing vibronic states is encoded in the frequency of this oscillation. To visualize the dynamical evolution of this frequency, the signal trace $S(t)$, be it the c-TRUECARS signal or the s-TRUECARS correlation function, is convolved with a Gaussian gating function $E_{\mathrm{gate}}(t)$ with 3~fs FWHM, scanning the trace at each time delay $T_{\mathrm{coh}}$, similar to a frequency-resolved optical-gating (FROG) measurement \cite{Trebino1997}: 

\begin{equation}
I_{\mathrm{FROG}}(\omega_{\mathrm{coh}},T_{\mathrm{coh}}) = \left| \int_{-\infty}^{\infty}  \diff t \,S(t) E_{\mathrm{gate}} (t-T_{\mathrm{coh}})\, \eu^{-\uimm \omega_{\mathrm{coh}} t}  \right|^2.
\label{eq:frog}
\end{equation}

\begin{figure}[t]
	\noindent\begin{centering}
		\includegraphics[width=\linewidth]{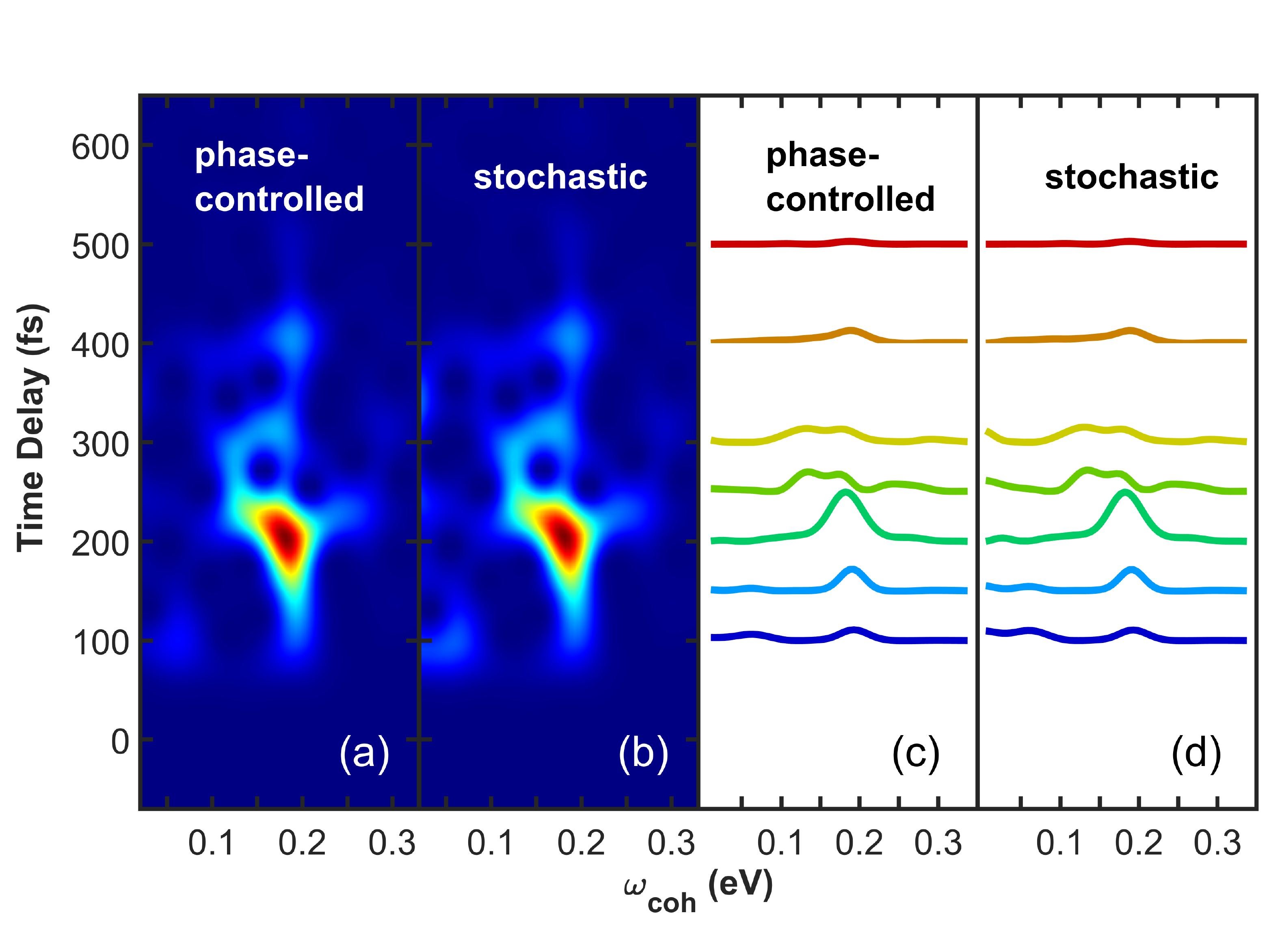}
		\par\end{centering}
	\caption{FROG spectrograms of the signal traces shown in Fig.~\ref{fig:slices}(a) according to Eq.~(\ref{eq:frog}). When the nuclear WP reaches the CoIn at 100~fs, the S\textsubscript{2}/S\textsubscript{1}  coherence emerges at 0.2~eV, from where it evolves to lower energies at 250~fs. The spectrogram of (a) the \ctc signal trace and (b) the \stc trace both reveal this information. Representative vertical slices of the (c) \ctc and (d) \stc spectrograms at indicated times corroborate this similarity.}
	\label{fig:frog}
\end{figure}

This yields a spectrogram of the signal trace $S(t)$ which reveals the energy splitting of the coherence $\omega_{\mathrm{coh}}$ at each delay. The spectrograms for both the phase-controlled and the stochastic signal are shown in Fig.~\ref{fig:frog}. Both the spectrograms in Figs.~\ref{fig:frog}(a) and \ref{fig:frog}(b), as well as the representative slices at indicated time delays in Figs.~\ref{fig:frog}(c) and \ref{fig:frog}(d), are very similar. The coherence phase evolves from higher values of 0.2~eV at 100~fs to lower values at 250~fs, mapping the evolution of the WP coherence on the electronic potential energy surface around the CoIn. Strikingly, this information is accessible in equivalent detail using stochastic pulses.

\begin{figure}[t]
	\noindent\begin{centering}
		\includegraphics[width=\linewidth]{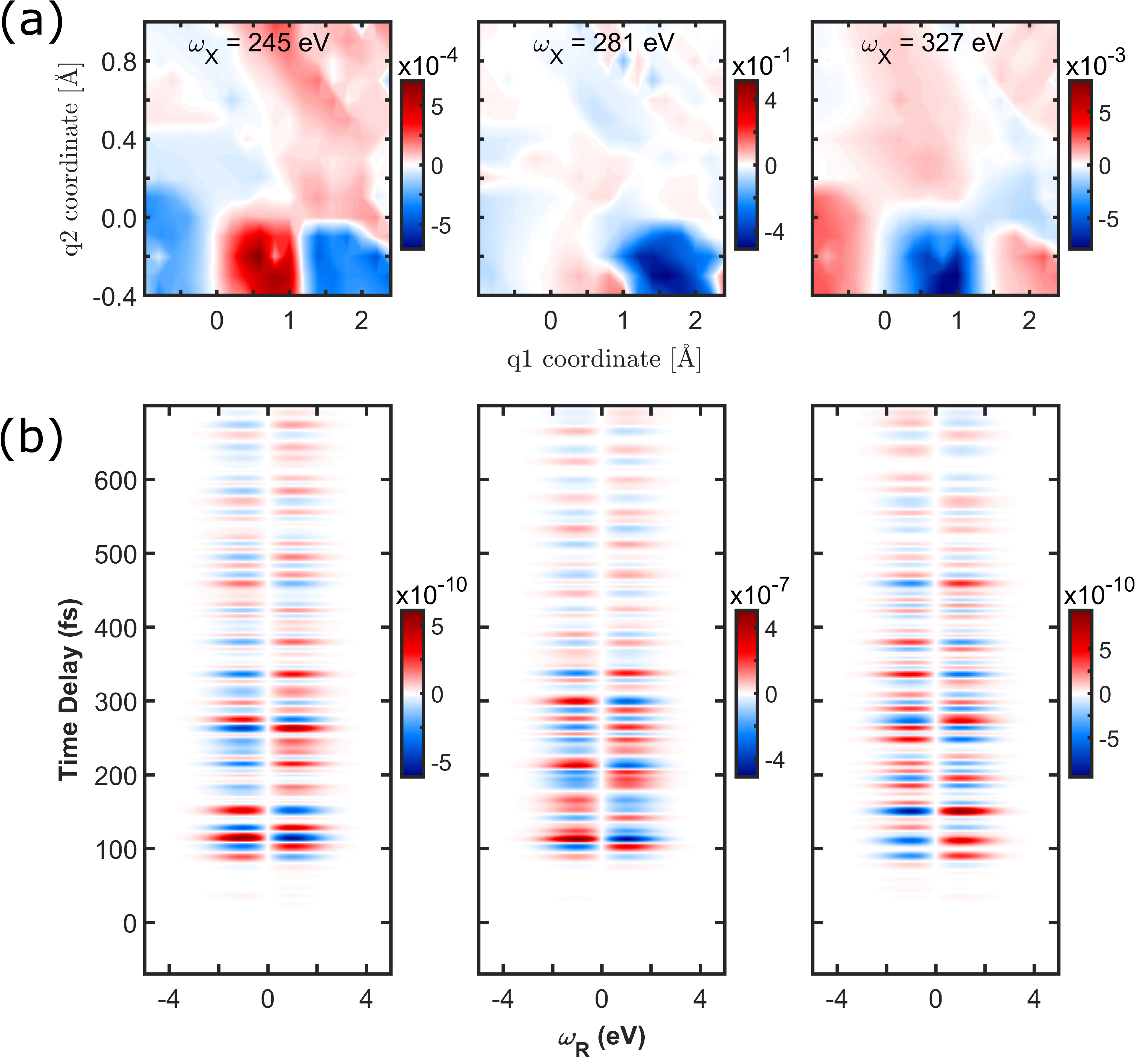}
		\par\end{centering}
	\caption{Molecular polarizabilities of Uracil and s-TRUECARS signal at three different probe wavelengths for $z$-polarized x-ray pulses. (a) Polarizability from Eq.~(\ref{eq:polarizability}) in atomic units, in the two-dimensional nuclear space of the Uracil Hamiltonian \cite{Keefer2017,Keefer2020}. $q_1$ and $q_2$ are the nuclear degrees of freedom, with $q_1$ leading from the Franck--Condon point to the CoIn, and $q_2$ leading to a local S\textsubscript{2} minimum. The indicated probe frequency enters Eq.~(\ref{eq:polarizability}) as $\omega_{\mathrm{X}}$, with the middle panel at 281~eV, 10~eV below the bound state Carbon resonance, and the left and right panels significantly below and above this resonance, respectively. (b) s-TRUECARS signal associated with these polarizabilities. The same qualitative features are exhibited, with the signal in the middle panel being three orders of magnitude stronger than the other two.}
	\label{fig:strengths}
\end{figure}

The magnitude of the molecular polarizability determines the strength of the s-TRUECARS signal, and thus its ability to survive loss processes. In Fig.~\ref{fig:strengths}, we display the polarizability for Uracil in the nuclear space of the two reactive coordinates in the effective Hamiltonian~\cite{Keefer2020}. These were calculated according to Eq.~(\ref{eq:polarizability}), and are dependent on the probe pulse carrier frequency $\omega_{\mathrm{X}}$. Three cases are shown, with $\omega_{\mathrm{X}}=245\,\mathrm{eV}$, below the Carbon resonance, $\omega_{\mathrm{X}} = 281\,\mathrm{eV}$, 10~eV below the Carbon resonance, and $\omega_{\mathrm{X}} = 327\,\mathrm{eV}$, between the Carbon and the Nitrogen resonance. When close to a bound state resonance, the polarizability becomes significantly stronger, in this case by around three orders of magnitude. The signal is visible in all three cases, with the same qualitative features, but is also enhanced by three orders of magnitude for $\omega_{\mathrm{X}}$ closer to the Carbon resonance. This shows that even within the parameter space determined by a given molecule, the polarizability, and thus the strength of the s-TRUECARS signal compared to other competing processes, can be tuned heavily. Also in other molecules, with different, weaker or stronger polarizabilities, the x-ray pulse frequency could be used to control the s-TRUECARS signal strength.

\section{Conclusions and outlook}
\label{Sec:Conclusions}
The stochastic properties of x-ray FEL pulses are commonly assumed to have a detrimental effect on the joint temporal and spectral resolution of spectroscopic signals compared to coherent pulses. Intense phase-controlled pulses which can be reproduced from shot to shot are not yet available at hard-x-ray FELs based on the SASE mechanism. We have shown that, by taking advantage of the correlations of the field, stochastic FEL pulses can provide the same temporal and spectral resolution as phase-controlled pulses. Like its coherent counterpart, s-TRUECARS offers a probe of the evolution of molecular coherences free from the background owing to the populations. The signal arises from the interaction with a single stochastic pulse, without requiring any phase control, and information is retrieved by averaging over many independent repetitions. Each signal originates from specific spectral components of the field, and time- and frequency-resolved spectroscopic information is extracted by exploiting the field correlations. 

In s-TRUECARS, the duration of the time envelope determines the time--frequency resolution, while the broad frequency envelope of the x-ray FEL pulse offers a large observation bandwidth and a well defined central time $T$. UDSP pulses with $a<\pi$ were shown to provide this combination of properties, as well as the stochastic pulses discussed in Appendix~\ref{Appendix:C+S}, consisting of the sum of a short peaked pulse and a background broadband noise. The latter scheme could be experimentally realized with attosecond pulses recently demonstrated at FELs \cite{duris2020tunable, maroju2020attosecond}. No control over the phase of the pulse is necessary, and shot-to-shot variations of the FEL pulse do not hinder the application of the technique. In contrast, UDSP pulses with $a=\pi$, similar to models employed in Refs.~\cite{vannucci1980computer, pfeifer2010partial, kimberg2016stochastic}, were shown to lead to the same limitations in joint spectral and temporal resolutions as c-TRUECARS implemented with a single pulse.

Different stochastic-field models and statistics, as enabled by recent advances in the shaping of, e.g., XUV FEL pulses \cite{gauthier2015spectrotemporal}, could be considered. In the UDSP model, a finite correlation frequency emerges via the gate function $\tilde{u}(\omega)$. In Appendix~\ref{App:cumulant}, we present a different stochastic-phase model with Gaussian statistics, which can be calculated exactly by the second-order cumulant expansion.

The s-TRUECARS signal of Eq.~(\ref{eq:correlationearlydefinition}) is defined by the correlation function between the frequency-resolved signal and the spectral intensity of the incoming pulse. Alternatively, one could correlate the spectral intensity of the transmitted, outgoing x-ray pulse at different frequencies. For optically thin samples, the signal in Eq.~(\ref{eq:signaldefinition}) is given by the difference between the outgoing and the incoming spectral intensities. Correlating the spectra measured after transmission through the sample will then yield a function corresponding to the correlation function in Eq.~(\ref{eq:correlationearlydefinition}) added to the autocorrelation of the incoming x-ray pulse. While this could render the analysis and extraction of spectrally and temporally resolved information more challenging, it would simplify the experimental implementation of the technique.

The present approach can be extended to other nonlinear signals, with any number of interactions with the stochastic field. TRUECARS implemented with stochastic x-ray pulses resonantly tuned to the core-state transitions in the molecule is a straightforward extension of the off-resonant case described here. A recent investigation in Thiophenol molecules with phase-controlled pulses showed that resonant TRUECARS offers temporally and spectrally resolved information about the dynamics of the molecular wavepacket \cite{cho2020stimulated}, albeit with background contributions from the populations. Our approach, based on the correlations of stochastic fields and exemplified in Sec.~\ref{Sec:s-TRUECARS} for off-resonant TRUECARS, could be straightforwardly applied to resonant x-ray pulses as well, with a significant increase in the molecular polarizability and the associated signal strength. More in general, the methods implemented here for the calculation of the two- and four-point correlation functions can be applied to derive higher-order $n$-point correlation functions for the prediction of signals involving $n$ stochastic fields. Such signals could involve correlations obtained by post-processing of the data, as was the case here, but could also represent the direct outcome of other measurements. This will allow the extension of virtually any multidimensional spectroscopy protocols from the optical to the hard-x-ray regime at present-day FEL facilities, with promising applications to the study and control of ultrafast electronic dynamics in complex molecular systems with light and, beyond that, in proteins or semiconductors.

\begin{acknowledgments}
The support of the Chemical Sciences, Geosciences, and Biosciences division, Office of Basic Energy Sciences, Office of Science, U.S. Department of Energy through Award DE-FG02-04ER15571 and of the National Science Foundation (Grant CHE-1953045) is gratefully acknowledged. S.M.C. and D.K. were partially supported by the DOE grant. S.M.C. and D.K. gratefully acknowledge the support of the Alexander von Humboldt foundation through the Feodor~Lynen program. We thank Nora~Berrah and Robert~W.~Schoenlein for most valuable discussions.
\end{acknowledgments}

\appendix

\section{Two- and four-point correlation functions of stochastic UDSP pulses}
\label{App:uniformdist}

In this Appendix, we derive the two- and four-point correlation functions of the pulses in Eq.~(\ref{eq:stochpulse}). Since the chaotic nature of the pulses stems from the stochastic phase $\varphi(\omega)$, we start by considering the two- and four-point correlation functions of $\eu^{\uimm\varphi(\omega)}$, which can be calculated exactly if $\varphi(x)$ and $\varphi(y)$, $x\neq y$, are independent random variables. For the UDSP model, as given in Eq.~(\ref{eq:stochphase}), the two- and four-point correlation functions of $\eu^{\uimm\varphi(\omega)}$ read 
\begin{equation}
\begin{aligned}
\tilde{F}_2(x, y)  \,&\doteq \langle\eu^{-\uimm\varphi(x)}\,\eu^{\uimm\varphi(y)} \rangle\\
&=s(a)^2+\varLambda\,\left[1-s^2(a)\right]\,\delta(x - y)
\end{aligned}
\label{eq:tildeF2uniformstat}
\end{equation}
and
\begin{widetext}
\begin{equation}
\begin{aligned}
\tilde{F}_4(x, y, x', y') \doteq\,& \langle\eu^{-\uimm\varphi(x)}\,\eu^{\uimm\varphi(y)}\,\eu^{-\uimm\varphi(x')}\,\eu^{\uimm\varphi(y')} \rangle\\
=\,&s^4(a) + \varLambda\,s^2(a) \Bigl\{\left[1-s^2(a)\right]\Bigl(\delta(x-y)+\delta(x-y') + \delta(y-x') +\delta(x'-y') \Bigr) \\
&\ \ \ \ \ \ \ \ \ -s(a)\,[s(a) - c(a)]\Bigl(\delta(x-x') + \delta(y-y') \Bigr)\Bigr\} \\
&+ \varLambda^2\Bigl\{ \left[1-s^2(a)\right]^2\Bigl(\delta(x-y)\delta(x'-y') +\delta(x-y')\delta(y-x')\Bigr) \\
&\ \ \ \ \ \ \ \ \ - s^2(a)\,\left[1 - 2s^2(a) + s(a)\,c(a)\right] \Bigl(\delta(x-y) + \delta(x'-y')\Bigr)\,\Bigl(\delta(x-y')+ \delta(y-x')\Bigr)\\
&\ \ \ \ \ \ \ \ \ \ \ \ \ \ \ \ \ \ +  s^2(a)\,[s(a)-c(a)]^2\,\delta(x-x')\delta(y-y')\Bigr\} \\
&-\varLambda^3\,\left\{1- s^2(a)\,\left[4-c^2(a)\right] - c(a)\,s^3(a) - 6s^4(a)\right\}\,\delta(x-y)\delta(x-x')\delta(x-y'),
\end{aligned}
\label{eq:tildeF4uniformstat}
\end{equation}
\end{widetext}
respectively, where $s(a) = \sinc(a)$ and $c(a) = \cos(a)$, and where we have substituted Kronecker deltas, which apply to discrete independent phases $\varphi_i$ with Dirac delta functions, modeling uncorrelated continuous phases $\varphi(\omega)$:
\begin{equation}
\delta_{ij} \xrightarrow[\varLambda \tau \ll 1]{} \varLambda\,\delta(\omega - \omega').
\end{equation}
This is a valid substitution for $\varLambda\ll 1/\tau$, as required in order to reproduce the spiky frequency envelopes of experimental FEL pulses. The two- and four-point correlation functions in Eqs.~(\ref{eq:tildeF2uniformstat}) and (\ref{eq:tildeF4uniformstat}) consist of a sum of products of delta functions, with different contributions reflecting whether any of the two (four) frequencies in $\tilde{F}_2(x,y)$ [$\tilde{F}_4(x,y,x',y')$] are identical. The coefficients in front of each addend were calculated via the probability density function $P(\varphi)$ in Eq.~(\ref{eq:stochphase}). An alternative, approximate approach, based on the second-order cumulant expansion and exact only for Gaussian statistics, is presented in Appendix~\ref{App:cumulant}. 

The two- and four-point correlation functions of $\tilde{E}(\omega)$ are given by
\begin{equation}
\begin{aligned}
&F_2(\omega_1, \omega_2) = \langle\tilde{E}^*(\omega_1)\tilde{E}(\omega_2)\rangle \\
=\,& \int \diff x \int \diff y \,\tilde{g}^*(x)\,\tilde{g}(y)\,\tilde{F}_2(x, y)\,\tilde{u}(\omega_1-x)\,\tilde{u}(\omega_2-y)
\end{aligned}
\end{equation}
and
\begin{equation}
\begin{aligned}
&F_4(\omega_1, \omega_2, \omega_3, \omega_4) = \langle\tilde{E}^*(\omega_1)\tilde{E}(\omega_2)\tilde{E}^*(\omega_3)\tilde{E}(\omega_4)\rangle \\
=\,& \int \diff x \int \diff y\int \diff x'\int \diff y'\,\tilde{g}^*(x)\,\tilde{g}(y)\,\tilde{g}^*(x')\,\tilde{g}(y')\\
&\ \ \ \ \ \ \times \tilde{F}_4(x, y, x', y')\, \tilde{u}(\omega_1-x)\,\tilde{u}(\omega_2-y)\\
&\ \ \ \ \ \ \ \ \ \ \ \ \times\tilde{u}(\omega_3-x')\,\tilde{u}(\omega_4-y'),
\end{aligned}
\end{equation}
with $\tilde{F}_{2}(x,y)$ and $\tilde{F}_{4}(x,y,x',y')$ from Eqs.~(\ref{eq:tildeF2uniformstat}) and (\ref{eq:tildeF4uniformstat}), respectively. By assuming a broadband envelope function $\tilde{g}(\omega)$, the two- and four-point correlation functions read
\begin{widetext}
\begin{equation}
\begin{aligned}
F_2(\omega_1, \omega_2) 
\approx
\tilde{g}^*(\omega_1)\,\tilde{g}(\omega_2)\,\left\{2\pi\,s^2(a) + \left[1-s^2(a)\right]\,\varLambda\sqrt{\pi}\tau\,\eu^{-\tfrac{(\omega_1 - \omega_2)^2\tau^2}{4}}\right\}
\end{aligned}
\label{eq:F2uniformstat}
\end{equation}
and
\begin{equation}
\begin{aligned}
&F_4(\omega_1, \omega_2, \omega_3, \omega_4) \\
\approx\,& \tilde{g}^*(\omega_1)\,\tilde{g}(\omega_2)\,\tilde{g}^*(\omega_3)\,\tilde{g}(\omega_4)\biggl\{(2\pi)^2\,s^4(a)+ 2\pi\,s^2(a)\,\varLambda\sqrt{\pi}\tau\,\biggl[\left[1- s^2(a)\right]\biggl(\eu^{-\tfrac{(\omega_1 - \omega_2)^2\tau^2}{4}}+\eu^{-\tfrac{(\omega_1 - \omega_4)^2\tau^2}{4}}\\
&\ \ \ \ \ \ +\eu^{-\tfrac{(\omega_2 - \omega_3)^2\tau^2}{4}}+\eu^{-\tfrac{(\omega_3 - \omega_4)^2\tau^2}{4}}\biggr) - s(a)[s(a) - c(a)]\,\biggl(\eu^{-\tfrac{(\omega_1 - \omega_3)^2\tau^2}{4}}+\eu^{-\tfrac{(\omega_2 - \omega_4)^2\tau^2}{4}} \biggr)\biggr]\\
&\ \ \ \ \ \ \ \ \ \ \ \ + \left[1-s^2(a)\right]^2 \varLambda^2\pi\tau^2\Bigl(\eu^{-\tfrac{(\omega_1-\omega_2)^2\tau^2}{4}}\,\eu^{-\tfrac{(\omega_3-\omega_4)^2\tau^2}{4}} + \eu^{-\tfrac{(\omega_1-\omega_4)^2\tau^2}{4}}\,\eu^{-\tfrac{(\omega_2-\omega_3)^2\tau^2}{4}}\Bigr)+\cdots\biggr\}.
\end{aligned}
\label{eq:F4uniformstat}
\end{equation}
\end{widetext}

\section{Energy-jitter effects}
\label{App:jitter}

To account for shot-to-shot variations in the pulse central frequency $\omega_{\mathrm{X}}$ due to machine drifts at FELs, a small frequency shift $\epsilon$, varying from shot to shot, can be included in the envelope of each stochastic pulse,
\begin{equation}
E_{\epsilon}(t) = 2\pi\,f(t)\,u(t)\,\eu^{-\uimm\epsilon t},
\end{equation}
with the same definitions of Sec.~\ref{Sec:pulses} and the resulting electric field given by Eq.~(\ref{eq:classicalfield}). This is associated with the spectral envelope
\begin{equation}
\begin{aligned}
\tilde{E}_{\epsilon}(\omega) =\,&  \int\diff t\,E_{\epsilon}(t)\,\eu^{\uimm\omega t}\\
=\,& \int \diff \omega'\,\tilde{g}(\omega'-\epsilon)\,\eu^{\uimm\varphi(\omega' - \epsilon)}\,\tilde{u}(\omega - \omega')
\end{aligned}
\end{equation}
and the two- and four-point correlation functions
\begin{equation}
\begin{aligned}
&F_{2,\epsilon}(\omega_1, \omega_2)  = \langle\tilde{E}_{\epsilon}^*(\omega_1)\tilde{E}_{\epsilon}(\omega_2)\rangle \\
=\,&  \int \diff x \int \diff y \,\tilde{g}^*(x-\epsilon)\,\tilde{g}(y-\epsilon)\,\tilde{F}_2(x, y)\,\tilde{u}(\omega_1-x)\,\tilde{u}(\omega_2-y)
\end{aligned}
\end{equation}
and
\begin{equation}
\begin{aligned}
&F_{4,\epsilon}(\omega_1, \omega_2, \omega_3, \omega_4) = \langle\tilde{E}_{\epsilon}^*(\omega_1)\tilde{E}_{\epsilon}(\omega_2)\tilde{E}_{\epsilon}^*(\omega_3)\tilde{E}_{\epsilon}(\omega_4)\rangle \\
=\,& \int \diff x \int \diff y\int \diff x'\int \diff y'\,\tilde{g}^*(x-\epsilon)\,\tilde{g}(y-\epsilon)\,\tilde{g}^*(x'-\epsilon)\,\tilde{g}(y'-\epsilon)\\
&\ \ \ \ \ \ \times \tilde{F}_4(x, y, x', y')\, \tilde{u}(\omega_1-x)\,\tilde{u}(\omega_2-y)\\
&\ \ \ \ \ \ \ \ \ \ \ \ \times\tilde{u}(\omega_3-x')\,\tilde{u}(\omega_4-y'),
\end{aligned}
\end{equation}
where $\tilde{F}_{2}(x,y)$ and $\tilde{F}_{4}(x,y,x',y')$ are independent of the shift $\epsilon$, according to their definitions in Eqs.~(\ref{eq:tildeF2uniformstat}) and (\ref{eq:tildeF4uniformstat}). For broadband stochastic pulses, whose bandwidths are larger than the shift caused by machine drifts, this effect will induce a minor modulation in the strength of the signal, and will not modify the spectral properties of the s-TRUECARS signal discussed in Sec.~\ref{Sec:s-TRUECARS}.

\section{Alternative stochastic-pulse scheme}
\label{Appendix:C+S}

Here, we put forward an alternative stochastic-pulse scheme resulting from the sum of a short broadband pulse $\tilde{E}_{\mathrm{c}}(\omega) = \sqrt{2\pi}s\,\tilde{g}(\omega)$ and a stochastic UDSP FEL pulse $\tilde{E}_{\text{$\pi$-UDSP}}(\omega)$ with $a = \pi$,
\begin{equation}
\begin{aligned}
&\tilde{E}(\omega) = \tilde{E}_{\mathrm{c}}(\omega) + \tilde{E}_{\text{$\pi$-UDSP}}(\omega)\\
=\,&\sqrt{2\pi}s\,\tilde{g}(\omega) + \sqrt{1-s^2}\,\tilde{g}(\omega)\int\diff\omega'\,\eu^{\uimm\varphi_{\pi}(\omega')}\,\tilde{u}(\omega - \omega'),
\end{aligned}
\label{eq:sumpulse}
\end{equation}
where $\varphi_{\pi}(\omega)$ is a UDSP function varying in $[-\pi,\,\pi]$. Note that neither shaping is required, not control over the absolute or relative phases of $\tilde{E}_{\mathrm{c}}(\omega)$ and $\tilde{E}_{\text{$\pi$-UDSP}}(\omega)$, and the pulse could thus be obtained at x-ray FELs.

The two- and four-point correlation functions of the stochastic pulse $\tilde{E}(\omega)$ in Eq.~(\ref{eq:sumpulse}), given by
\begin{widetext}
\begin{equation}
F_2(\omega_1, \omega_2) \doteq \langle\tilde{E}^*(\omega_1)\tilde{E}(\omega_2)\rangle  \\
=
\tilde{g}^*(\omega_1)\,\tilde{g}(\omega_2)\,\left(2\pi\,s^2 + \left(1-s^2\right)\,\varLambda\sqrt{\pi}\tau\,\eu^{-\tfrac{(\omega_1 - \omega_2)^2\tau^2}{4}}\right)
\label{eq:F2sumpulse}
\end{equation}
and
\begin{equation}
\begin{aligned}
&F_4(\omega_1, \omega_2, \omega_3, \omega_4) \doteq \langle\tilde{E}^*(\omega_1)\tilde{E}(\omega_2)\tilde{E}^*(\omega_3)\tilde{E}(\omega_4)\rangle \\
\approx\,& \tilde{g}^*(\omega_1)\,\tilde{g}(\omega_2)\,\tilde{g}^*(\omega_3)\,\tilde{g}(\omega_4)\biggl[(2\pi)^2\,s^4+ 2\pi\,s^2\,(1-s^2)\,\varLambda\sqrt{\pi}\tau\,\biggl(\eu^{-\tfrac{(\omega_1 - \omega_2)^2\tau^2}{4}}+\eu^{-\tfrac{(\omega_1 - \omega_4)^2\tau^2}{4}}+\eu^{-\tfrac{(\omega_2 - \omega_3)^2\tau^2}{4}}\\
&\ \ \ +\eu^{-\tfrac{(\omega_3 - \omega_4)^2\tau^2}{4}}\biggr)+ \left(1-s^2\right)^2\, \varLambda^2\pi\tau^2\Bigl(\eu^{-\tfrac{(\omega_1-\omega_2)^2\tau^2}{4}}\,\eu^{-\tfrac{(\omega_3-\omega_4)^2\tau^2}{4}} + \eu^{-\tfrac{(\omega_1-\omega_4)^2\tau^2}{4}}\,\eu^{-\tfrac{(\omega_2-\omega_3)^2\tau^2}{4}}\Bigr)+\cdots \biggr],
\end{aligned}
\label{eq:F4sumpulse}
\end{equation}
\end{widetext}
and those of a UDSP pulse with $a<\pi$, shown in Eqs.~(\ref{eq:F2uniformstat}) and (\ref{eq:F4uniformstat}), are closely related. For $s = s(a)$, the two-point correlation functions~(\ref{eq:F2uniformstat}) and (\ref{eq:F2sumpulse}) are identical, and the four-point correlation functions~(\ref{eq:F4uniformstat}) and (\ref{eq:F4sumpulse}) display an analogous dependence upon the Gaussian envelope $\tilde{u}(\omega/\sqrt{2}) = \tau\,\eu^{-\omega^2\tau^2/4}$. 

These analogous two- and four-point correlation functions lead to identical s-TRUECARS signals. By inserting the pulse two- and four-point correlation functions~(\ref{eq:F2sumpulse}) and (\ref{eq:F4sumpulse}) into Eq.~(\ref{eq:correlationdefinition}), the correlation function to first order in $(\varLambda \tau)$ reads
\begin{widetext}
\begin{equation}
\begin{aligned}
C(\omega_{\mathrm{R}} ,\omega_{\mathrm{m}} , T) \,&=4\pi s^2\left(1-s^2\right)\varLambda\Imm\biggl\{ \Bigl|\tilde{g}\Bigl(\omega_{\mathrm{m}}-\frac{\omega_{\mathrm{R}}}{2}\Bigr)\Bigr|^2\,\tilde{g}^*\Bigl(\omega_{\mathrm{m}}+\frac{\omega_{\mathrm{R}}}{2}\Bigr)\\
&\ \ \ \times \int\frac{ \diff\omega}{2\pi} \,\tilde{g}\Bigl(\omega_{\mathrm{m}}+\frac{\omega_{\mathrm{R}}}{2}-\omega\Bigr)\,\sqrt{\pi}\tau\,\biggl(\eu^{-\tfrac{\omega_{\mathrm{R}2}^2\tau^2}{4}} + \eu^{-\tfrac{(\omega_{\mathrm{R}}- \omega)^2\tau^2}{4}}\biggr)\,\eu^{-\uimm\omega T}\,\langle\hat{\tilde{\alpha}}(\omega)\rangle \biggr\},
\end{aligned}
\label{eq:correlationfunctionsumpulse}
\end{equation}
\end{widetext}
with the same structure as the s-TRUECARS signal in Eq.~(\ref{eq:correlationfunction}) for UDSP pulses with $a<\pi$.

\section{Derivation of effective polarizability and TRUECARS signal via the minimal-coupling light--matter interaction Hamiltonian}
\label{App:minimalcoupling}

The minimal-coupling Hamiltonian provides the complete formalism to describe the interaction between light and matter by avoiding the multipolar expansion. In the rotating-wave approximation, this is given by
\begin{equation}
\begin{aligned}
\hat{H}_{\mathrm{int}} &= -\int \diff^3 r\biggl(\hat{j}^{(+)}(\boldsymbol{r}) \hat{\mathcal{A}}(\boldsymbol{r}) + \hat{j}^{(-)}(\boldsymbol{r}) \hat{\mathcal{A}}\daga(\boldsymbol{r}) \\
&\ \ \ \ \ \ \ \ - \frac{1}{2}\hat{\sigma}(\boldsymbol{r})\hat{\mathcal{A}}\daga(\boldsymbol{r})\hat{\mathcal{A}}(\boldsymbol{r})\biggr).
\end{aligned}
\label{eq:intHam-minimalcoupling}
\end{equation}
Here, the matter is described in terms of the charge-density operator $\hat{\sigma}(\boldsymbol{r})$ and the positive- and negative-frequency parts of the current-density operator $\hat{j}(\boldsymbol{r}) = \hat{j}^{(+)}(\boldsymbol{r}) + \hat{j}^{(-)}(\boldsymbol{r})$. The radiation field is given by the vector-potential operator 
\begin{equation}
\hat{\mathcal{A}}(\boldsymbol{r}) = \sum_j \sqrt{\frac{2\pi}{V\omega_j}}\,\hat{a}(\omega_j)\,\eu^{\uimm\boldsymbol{k}_j \cdot\boldsymbol{r}}
\end{equation}
via the radiation modes $\boldsymbol{k}_j$. In the above equations, $\hat{\mathcal{A}}(\boldsymbol{r})$ and $\hat{j}(\boldsymbol{r})$ are the projections of the associated vectors along a fixed field-polarization direction. 

With steps analogous to those used in the derivation of Eq.~(\ref{eq:signal}), the signal is obtained via Heisenberg equations of motion for the photon number operator, assuming also here that the x-ray radiation is in a coherent state such that the vector-potential operator $\hat{\mathcal{A}}(\boldsymbol{r})$ can be replaced by the classical field
\begin{equation}
\mathcal{A}(\boldsymbol{r}, t) = A(t - T)\,\eu^{\uimm\boldsymbol{k}_{\mathrm{X}}\cdot\boldsymbol{r}}\,\eu^{-\uimm\omega_{\mathrm{X}}(t-T)},
\end{equation}
with the complex envelope functions $A(t)$, wavevector $\boldsymbol{k}_{\mathrm{X}}$, and frequency-domain envelope $\tilde{A}(\omega) = \int \diff t\, A(t)\,\eu^{\uimm\omega t}$. The frequency-dispersed spectrum $S(\omega_{\mathrm{s}})$ as a function of the pulse arrival time $T$ is then given by
\begin{equation}
\begin{aligned}
&S(\omega_{\mathrm{s}}, T)\\
=\,& -2\Imm\biggl\{\tilde{A}^*(\omega_{\mathrm{s}} - \omega_{\mathrm{X}})  \int \diff t\,A(t-T)\,\eu^{\uimm(\omega_{\mathrm{s}} - \omega_{\mathrm{X}})(t-T)}\\
&\times \Bigl\langle\biggl[\int\diff^3 r \int\diff^3 r'\,\sum_{c}\frac{\hat{j}^{(-)}(\boldsymbol{r},t)|c\rangle\langle c| \hat{j}^{(+)}(\boldsymbol{r'},t)}{\omega_{\mathrm{X}}- \omega_c + \uimm\gamma_c}\,\eu^{-\uimm\boldsymbol{k}_{\mathrm{X}}\cdot(\boldsymbol{r} - \boldsymbol{r}')}\\
&\ \ \ \ \ \ \ +\frac{1}{2}\int\diff^3 r\, \hat{\sigma}(\boldsymbol{r},t) \biggr]\Bigr\rangle\biggr\}.
\label{eq:signal-minimalcoupling}
\end{aligned}
\end{equation}
The sum in $c$ runs over all possible high-energy excited states which are coupled to the molecular vibronic states by the x-ray pulse, and can thus include off-resonant bound core-excited states and continuum states. By substituting
\begin{equation}
E(t) = \uimm\,\omega_{\mathrm{X}}\,A(t)
\end{equation}
and defining the effective polarizability operator
\begin{equation}
\begin{aligned}
&\hat{\alpha}(t) \\
=- \,&\biggl[\int\diff^3 r \int\diff^3 r'\,\sum_{c}\frac{\hat{j}^{(-)}(\boldsymbol{r},t)|c\rangle\langle c| \hat{j}^{(+)}(\boldsymbol{r'},t)}{\omega_{\mathrm{X}}- \omega_c + \uimm\gamma_c}\,\eu^{-\uimm\boldsymbol{k}_{\mathrm{X}}\cdot(\boldsymbol{r}-\boldsymbol{r}')}\\
&\ \ +\frac{1}{2}\int\diff^3 r\, \hat{\sigma}(\boldsymbol{r},t)\biggr]\,\frac{1}{\omega_{\mathrm{X}}^2},
\end{aligned}
\end{equation}
Eq.~(\ref{eq:signal}) is recovered.

The polarizability operator employed in Sec.~\ref{Sec:s-TRUECARS-CoIn} for Uracil was calculated in Ref.~\cite{Keefer2020} from ab-initio theory in the dipole approximation,
\begin{equation}
\alpha_{ij} = \sum_c\biggl(\frac{\langle i|\hat{\mu}|c\rangle \langle c|\hat{\mu}|j\rangle}{\omega_{cj} - \omega_{\mathrm{X}}} + \frac{\langle i|\hat{\mu}|c\rangle \langle c|\hat{\mu}|j\rangle}{\omega_{ci} + \omega_{\mathrm{X}}} \biggr),
\label{eq:polarizability}
\end{equation}
where $\hat{\mu}$ is the component of the dipole-moment operator along the field polarization direction, and the second term in Eq.~(\ref{eq:polarizability}) includes contributions beyond the rotating-wave approximation. The sum in $c$ includes forty C, twenty N, and twenty O core-excited states. Coupling to the continuum was not included directly in the calculation of $\langle\hat{\alpha}(t)\rangle$. For off-resonant x-ray pulses, however, this direct pathway through continuum intermediate states was shown to induce population transfer \cite{nakajima1994population}. Including the coupling to the continuum could thus contribute to the effective polarizability $\langle\hat{\alpha}(t)\rangle$ between the vibronic states in the molecule, whose evolution is probed by the TRUECARS signal. This would not alter the definition of the TRUECARS signal in Eq.~(\ref{eq:signal}), but only the explicit form of $\langle\hat{\alpha}(t)\rangle$ therein. Nondipole effects, not included in Eq.~(\ref{eq:polarizability}), have been investigated for resonant and off-resonant x-ray spectroscopy \cite{tanaka2001time, rouxel2016current, cavaletto2020probing}, and were recently shown to affect molecular photoionization \cite{grundmann2020zeptosecond}. New simulations in Uracil based on the minimal-coupling interaction Hamiltonian may modify the details of the spectra in Figs.~\ref{fig:signal}--\ref{fig:strengths}, but would not alter the applicability of s-TRUECARS with stochastic FEL pulses.

\section{Influence of x-ray photoionization on the TRUECARS signal}
\label{App:Photoionization}

The TRUECARS signal in Eq.~(\ref{eq:signal}) provides access to the molecular dynamics via the polarizability $\langle\hat{\alpha}(t)\rangle$. In the most general case, this is obtained by solving the time-dependent Schr\"odinger equation for the molecule interacting with the x-ray probe fields. In particular, resonant coupling to the continuum can lead to photoionization and population losses at a rate 
\begin{equation}
\varGamma_{ij}(t) = \frac{1}{2}(\sigma_{\mathrm{X},i} + \sigma_{\mathrm{X},j})\,\mathcal{I}(t),
\label{eq:photoionization}
\end{equation}
where $\mathcal{I}(t) = |A\,E(t)|^2/(8\pi\alpha\omega_{\mathrm{X}})$ is the x-ray pulse flux and $\sigma_{\mathrm{X},i} = \sigma_i(\omega_{\mathrm{X}})$ are the photoionization cross sections evaluated at the pulse frequency $\omega_{\mathrm{X}}$.

As a population loss channel, x-ray photoionization does not modify the definition of the TRUECARS signal in Eq.~(\ref{eq:signal}). However, it can modulate the free evolution of the molecular polarizability by an exponentially decaying factor, centered around $T$ and with time-dependent decay rates given by $\varGamma_{ij}(t-T)$. Including the coupling to the continuum as a photoionization loss channel was shown to be important in recent studies of XUV stimulated Raman adiabatic passage via autoionizing states \cite{li2014population}. An exponential decay of $\langle\hat{\alpha}(t)\rangle$ will cause a decrease in its amplitude and thus a reduction in the strength of the TRUECARS signal. Furthermore, such exponential decay can act as an additional temporal gate function in the Fourier transform of Eq.~(\ref{eq:signal}). The rate of decay induced by photoionization needs to be small compared to the transition frequency $\omega_{ba}(t)$ of the system, to ensure that a sufficient number of oscillations are captured by the signal within the time window determined by x-ray photoionization. The x-ray pulse flux $\mathcal{I}(t)$ should thus be optimized, such that the decay rates in Eq.~(\ref{eq:photoionization}) do not compromise the frequency resolution provided by the TRUECARS technique.

Far from the strong-field regime, and especially for off-resonant x-ray pulses, additional interactions with the x-ray radiation beyond those included in Fig.~\ref{fig:Feynman} can be safely neglected. For strong, resonant x-ray fields, however, well beyond the range of intensities of interest here, additional interactions with the probe pulses may induce Rabi oscillations in the populations of the system \cite{kimberg2016stochastic}, which would be reflected in the spectral features of the signal.

\section{Estimation of the signal-to-background ratio}
\label{App:signal-to-background}

The signal-to-background ratio can be estimated by comparing the number of absorbed photons $N_{\mathrm{mol}}\,S(\omega_{\mathrm{s}})\,\diff \omega_{\mathrm{s}}$ to the number of probe-pulse photons $A_{\mathrm{foc}}\,\tilde{I}(\omega_{\mathrm{s}})/\omega_{\mathrm{X2}}\,\diff\omega_{\mathrm{s}}$ in the differential interval $\diff \omega_{\mathrm{s}}$ centered on $\omega_{\mathrm{s}}$. Here, $A_{\mathrm{foc}}$ is the focal area and
\begin{equation}
\tilde{I}(\omega_{\mathrm{s}}) = \frac{1}{8\pi\alpha}\,\frac{|\tilde{E}_{\mathrm{2}}(\omega_{\mathrm{s}} - \omega_{\mathrm{X2}})|^2}{2\pi}
\end{equation}
the pulse spectral intensity, with the fine-structure constant $\alpha$. With the signal defined in Eq.~(\ref{eq:signal}), the signal-to-background ratio is given by
\begin{equation}
\begin{aligned}
R(\omega_{\mathrm{s}},T) =\,& 32\pi^2\alpha n_{\mathrm{mol}} L \,\omega_{\mathrm{X2}}\\
&\times\Imm\biggl\{\int\frac{ \diff\omega}{2\pi}\,\frac{\tilde{E}_1(\omega_{\mathrm{s}}- \omega_{\mathrm{X}1}- \omega)}{\tilde{E}_2(\omega_{\mathrm{s}}- \omega_{\mathrm{X}2})}\,\eu^{-\uimm\omega T}\langle\hat{\tilde{\alpha}}(\omega)\rangle \biggr\},
\end{aligned}
\end{equation}
with the molecular density $n_{\mathrm{mol}}$ and the propagation length $L$. In the above equation, we considered the general c-TRUECARS setup requiring two pulses $\mathcal{E}_1(t)$ and $\mathcal{E}_2(t)$. In s-TRUECARS, the two pulses coincide.\\

For an estimation of the signal-to-background ratio, we approximate it to $R\approx 32\pi^2\alpha n_{\mathrm{mol}} L \omega_{\mathrm{X2}}\langle\hat{\alpha}(T)\rangle$, where we have assumed that $\mathcal{E}_1(t)$ and $\mathcal{E}_2(t)$ have the same peak intensity (as for s-TRUECARS) and neglected their spectral details. Figure~\ref{fig:signal} shows that $\langle\hat{\alpha}(T)\rangle \sim 10^{-7}\,\mathrm{a.u.}$ when $\omega_{\mathrm{X}} = 354\,\mathrm{eV}$. However, a two-order-of-magnitude increase in the value of the polarizability can be obtained by approaching the Carbon resonance, as shown in Fig.~\ref{fig:strengths}. We thus assume $\langle\hat{\alpha}(T)\rangle \sim 10^{-5}\,\mathrm{a.u.}$ and $\omega_{\mathrm{X2}}\sim300\,\mathrm{eV}$. For realistic values of the molecular density $n_{\mathrm{mol}} = 1.6\times 10^{19}\,\mathrm{cm^{-3}}$ \cite{weninger2013stimulatedPRL} and a short propagation length of $L = 1\,\mathrm{mm}$, we estimate a signal-to-background ratio of $R\sim 1\%$.

\section{Multi-point field correlation functions for Gaussian phase fluctuations}
\label{App:cumulant}

In this Appendix, we calculate the two- and four-point correlation functions of the stochastic pulse in Eqs.~(\ref{eq:stochpulset}) and (\ref{eq:stochpulse}) assuming a stochastic phase $\varphi(\omega)$ with Gaussian statistics. In this case, $\langle\varphi(\omega)\rangle$ and $\langle\varphi(\omega)\varphi(\omega')\rangle$ fully determine the higher momenta of the stochastic process. 

The stochastic nature of the pulse in Eqs.~(\ref{eq:stochpulset}) and (\ref{eq:stochpulse}) follows from the phase $\varphi(\omega)$. We thus start by considering the two- and four-point correlation functions of $\eu^{\uimm\varphi(\omega)}$, defined as
\begin{equation}
\tilde{F}_{2}(\omega_1, \omega_2) \doteq \langle \eu^{-\uimm\varphi(\omega_1)}\,\eu^{\uimm\varphi(\omega_2)}\rangle
\end{equation}
and 
\begin{equation}
\tilde{F}_{4}(\omega_1, \omega_2, \omega_3, \omega_4) \doteq \langle \eu^{-\uimm\varphi(\omega_1)}\,\eu^{\uimm\varphi(\omega_2)}\,\eu^{-\uimm\varphi(\omega_3)}\,\eu^{\uimm\varphi(\omega_4)}\rangle,
\end{equation}
respectively. Since $\tilde{F}_{2}(\omega_1, \omega_2) = \tilde{F}_{4}(\omega_1, \omega_2, \omega_3, \omega_3)$ for any $\omega_3$, we derive the four-point correlation function and obtain $\tilde{F}_{2}(\omega_1, \omega_2)$ as a particular case. 

$\tilde{F}_{4}(\omega_1, \omega_2, \omega_3, \omega_4)$ is calculated via the second-order cumulant expansion \cite{mukamel1995principles}, which is exact for Gaussian statistics and is based on the ansatz
\begin{equation}
\begin{aligned}
\tilde{F}_{\lambda,4}(\omega_1, \omega_2, \omega_3, \omega_4) = \exp{\left(\sum_{n=1}^{\infty}\lambda^n G_n(\omega_1, \omega_2, \omega_3, \omega_4)\right)}.
\end{aligned}
\label{eq:cumulant}
\end{equation}
Here, we defined 
\begin{equation}
\begin{aligned}
&\tilde{F}_{\lambda,4}(\omega_1, \omega_2, \omega_3, \omega_4)\\
\doteq\,&\biggl\langle\exp\biggl(-\uimm\lambda\biggl[\int_{\omega_0}^{\omega_1}\diff \omega\,\dot{\varphi}(\omega) -\int_{\omega_0}^{\omega_2}\diff \omega\,\dot{\varphi}(\omega)\\
&\ \ \ \ \ \ \ \ \ \ \ +\int_{\omega_0}^{\omega_3}\diff \omega\,\dot{\varphi}(\omega)-\int_{\omega_0}^{\omega_4}\diff \omega\,\dot{\varphi}(\omega)\biggr]\biggr)\biggr\rangle,
\end{aligned}
\end{equation}
with the first derivative $\dot{\varphi}(\omega)$ of the stochastic phase, while the functions $G_n(\omega_1, \omega_2, \omega_3, \omega_4)$ are determined via a second-order expansion in $\lambda$ of both sides in Eq.~(\ref{eq:cumulant}). The right-hand side of Eq.~(\ref{eq:cumulant}) reads
\begin{equation}
\begin{aligned}
&\exp{\left(\sum_{n=1}^{\infty}\lambda^n G_n(\omega_1, \omega_2, \omega_3, \omega_4)\right)} \\
=\,& 1 + \lambda G_1(\omega_1, \omega_2, \omega_3, \omega_4) \\
&+ \lambda^2\left[G_2(\omega_1, \omega_2, \omega_3, \omega_4) + \frac{1}{2}G_1(\omega_1, \omega_2, \omega_3, \omega_4)\right] + \ldots
\end{aligned}
\label{eq:right-sec-ord}
\end{equation}
The expansion of the left-hand side of Eq.~(\ref{eq:cumulant}) gives
\begin{equation}
\begin{aligned}
&\tilde{F}_{\lambda,4}(\omega_1, \omega_2, \omega_3, \omega_4)\\
=\,&\ 1 - \uimm \lambda \,\bigl[\langle \varphi(\omega_1)\rangle - \langle \varphi(\omega_2)\rangle + \langle \varphi(\omega_3)\rangle - \langle \varphi(\omega_4)\rangle \bigr] \\
&- \frac{\lambda^2}{2} \Bigl\{h(\omega_1, \omega_1) + h(\omega_2, \omega_2) + h(\omega_3, \omega_3) + h(\omega_4, \omega_4)\\
&\ \ \ \ \ \ \ \ - 2\bigl[h(\omega_1, \omega_2) - h(\omega_1, \omega_3) + h(\omega_1, \omega_4) \\
&\ \ \ \ \ \ \ \ \ \ + h(\omega_2, \omega_3) - h(\omega_2, \omega_4) + h(\omega_3, \omega_4)\bigr] \Bigr\} + \ldots,
\end{aligned}
\label{eq:sec-order}
\end{equation}
where we have defined
\begin{equation}
h(\omega_a, \omega_b) = \int_{\omega_0}^{\omega_a}\diff \omega \int_{\omega_0}^{\omega_b} \diff \omega' \langle\dot{\varphi}(\omega) \dot{\varphi}(\omega') \rangle.
\end{equation}
We assume that the expectation value of the phase vanishes for any frequency $\omega$
\begin{equation}
\langle \varphi(\omega)\rangle = 0
\end{equation} 
and define the functions
\begin{equation}
p(\omega_a, \omega_b) =  \int_{\min(\omega_a, \omega_b)}^{\max(\omega_a, \omega_b)} \diff \omega \int_{\min(\omega_a, \omega_b)}^{\max(\omega_a, \omega_b)}\diff\omega'\, \langle\dot{\varphi}(\omega) \dot{\varphi}(\omega') \rangle 
\end{equation}
and
\begin{equation}
\tilde{h}(\omega_a, \omega_b) =  \int_{\omega_0}^{\min(\omega_a, \omega_b)} \diff \omega \int_{\min(\omega_a, \omega_b)}^{\max(\omega_a, \omega_b)}\diff\omega'\, \langle\dot{\varphi}(\omega) \dot{\varphi}(\omega') \rangle,
\end{equation}
so that
\begin{equation}
\begin{aligned}
&h[\max(\omega_a, \omega_b), \max(\omega_a, \omega_b)] \\
=\,& h[\min(\omega_a, \omega_b), \min(\omega_a, \omega_b)] + 2 \tilde{h}(\omega_a, \omega_b) + p(\omega_a, \omega_b)
\end{aligned}
\end{equation}
and thus
\begin{equation}
\begin{aligned}
h(\omega_a, \omega_b) &\doteq h[\min(\omega_a, \omega_b), \min(\omega_a, \omega_b)] + \tilde{h}(\omega_a, \omega_b)\\
& = \frac{1}{2}\bigl[h(\omega_a,\omega_a) + h(\omega_b, \omega_b) - p(\omega_a, \omega_b)\bigr].
\end{aligned}
\end{equation}
With the above identities, Eq.~(\ref{eq:sec-order}) reduces to
\begin{equation}
\begin{aligned}
&\tilde{F}_{\lambda,4}(\omega_1, \omega_2, \omega_3, \omega_4)\\
=\,&\ 1 - \frac{\lambda^2}{2} \Bigl[p(\omega_1, \omega_2) - p(\omega_1, \omega_3) + p(\omega_1, \omega_4) \\
&\ \ \ \ + p(\omega_2, \omega_3) - p(\omega_2, \omega_4) + p(\omega_3, \omega_4)\Bigr] + \ldots,
\end{aligned}
\label{eq:left-sec-ord}
\end{equation}
and a comparison of Eqs.~(\ref{eq:left-sec-ord}) and (\ref{eq:right-sec-ord}) shows that
\begin{equation}
G_1(\omega_1, \omega_2, \omega_3, \omega_4) = 0
\end{equation}
and
\begin{equation}
\begin{aligned}
&G_2(\omega_1, \omega_2, \omega_3, \omega_4) \\
=\,& - \frac{1}{2} \Bigl[p(\omega_1, \omega_2) - p(\omega_1, \omega_3) + p(\omega_1, \omega_4) \\
&\ \ \ \ + p(\omega_2, \omega_3) - p(\omega_2, \omega_4) + p(\omega_3, \omega_4)\Bigr].
\end{aligned}
\end{equation}
By taking the second-order cumulant expansion, the four-point correlation function of $\eu^{\uimm\varphi(\omega)}$ is given by
\begin{equation}
\begin{aligned}
&\tilde{F}_{4}(\omega_1, \omega_2, \omega_3, \omega_4) \\
=\,& \exp\biggl\{-\frac{1}{2}\bigl[p(\omega_1 ,\omega_2) -p(\omega_1 , \omega_3)+p(\omega_1 , \omega_4)\\
&\ \ \ \ \ \ \ \ \ \ + p(\omega_2 , \omega_3)-p(\omega_2 , \omega_4)+p(\omega_3 , \omega_4)\bigr]\biggr\}.
\end{aligned}
\end{equation}
When the fluctuations of the stochastic spectral phase $\varphi(\omega)$ are given by a wide-sense stationary process so that
\begin{equation}
\langle\dot{\varphi}(\omega) \dot{\varphi}(\omega') \rangle  = \tilde{q}(\omega - \omega'),
\end{equation}
it follows that 
\begin{equation}
\begin{aligned}
p(\omega_a, \omega_b) \,&= \int_{\min(\omega_a, \omega_b)}^{\max(\omega_a, \omega_b)} \diff \omega \int_{\min(\omega_a, \omega_b)}^{\max(\omega_a, \omega_b)}\diff\omega'\, \tilde{q}(\omega - \omega') \\
&= \tilde{p}(\omega_a - \omega_b),
\end{aligned}
\end{equation}
with
\begin{equation}
\tilde{p}(\omega) = \int_0^{\omega}\diff x \int_0^{\omega}\diff y \,\langle \dot{\varphi}(x)\dot{\varphi}(y)\rangle.
\label{eq:ptilde}
\end{equation}
The four-point correlation function then reads
\begin{equation}
\begin{aligned}
&\tilde{F}_{4}(\omega_1, \omega_2, \omega_3, \omega_4)\\
=\,&\exp\biggl\{-\frac{1}{2}\bigl[\tilde{p}(\omega_1 - \omega_2) -\tilde{p}(\omega_1 - \omega_3)+\tilde{p}(\omega_1 - \omega_4)\\
&\ \ \ \ \ \ \ \ \ \ + \tilde{p}(\omega_2 - \omega_3)-\tilde{p}(\omega_2 - \omega_4)+\tilde{p}(\omega_3 - \omega_4)\bigr]\biggr\},
\end{aligned}
\label{eq:4pointphi}
\end{equation}
while the two-point correlation function is given by
\begin{equation}
\tilde{F}_{2}(\omega_1, \omega_2)= \tilde{F}_{4}(\omega_1, \omega_2, \omega_3, \omega_3) = \eu^{-\tfrac{1}{2}\tilde{p}(\omega_1- \omega_2)},
\label{eq:2pointphi}
\end{equation}
since $\tilde{p}(0) = 0$. 

For independent random phases, where
\begin{equation}
\begin{aligned}
\langle \varphi(\omega)\rangle &= 0,\\
\langle \varphi(\omega)\varphi(\omega')\rangle &= \langle\varphi^2\rangle\,\varLambda\,\delta(\omega-\omega'),
\end{aligned}
\end{equation}
and thus
\begin{equation}
\langle\dot{\varphi}(\omega)\dot{\varphi}(\omega')\rangle = -\langle\varphi^2\rangle\varLambda\,\ddot{\delta}(\omega - \omega'),
\end{equation}
the integral of the correlation function of $\dot{\varphi}(\omega)$ tends to
\begin{equation}
\tilde{p}(\omega) = 2\langle\varphi^2\rangle[1-\varLambda\,\delta(\omega)],
\end{equation}
whose exponential is equal to
\begin{equation}
\eu^{\pm\frac{1}{2}\tilde{p}(\omega)} = \eu^{\pm\langle\varphi^2\rangle}+\left(1-\eu^{\pm\langle\varphi^2\rangle}\right)\varLambda\,\delta(\omega).
\end{equation}
Under those conditions, the two- and four-point correlation functions of $\eu^{\uimm\varphi(\omega)}$ read
\begin{equation}
\tilde{F}_2(x, y) = \eu^{-\langle\varphi^2\rangle}+\varLambda\,\left(1-\eu^{-\langle\varphi^2\rangle}\right)\delta(x-y)
\label{eq:tildeF2Gaussianstat}
\end{equation}
and
\begin{widetext}
\begin{equation}
\begin{aligned}
\tilde{F}_4(x, y, x', y') = \,& \eu^{-2\langle\varphi^2\rangle} + \varLambda\,\eu^{-\langle\varphi^2\rangle}\left(1- \eu^{-\langle\varphi^2\rangle}\right)\Bigl[\Bigl(\delta(x-y)+\delta(x-y') + \delta(y-x') +\delta(x'-y') \Bigr) \\
&- \eu^{-\langle\varphi^2\rangle}\Bigl(\delta(x-x') + \delta(y-y') \Bigr)\Bigr] + \varLambda^2\,\left(1- \eu^{-\langle\varphi^2\rangle}\right)^2\Bigl[\delta(x-y)\delta(x'-y') +\delta(x-y')\delta(y-x') \\
&- \eu^{-\langle\varphi^2\rangle}\Bigl(\delta(x-y) + \delta(x'-y')\Bigr)\,\Bigl(\delta(x-y')+ \delta(y-x')\Bigr) + \eu^{-2\langle\varphi^2\rangle}\delta(x-x')\delta(y-y')\Bigr] \\
&-\varLambda^3\,\left(1- \eu^{-\langle\varphi^2\rangle}\right)^4 \delta(x-y)\delta(x-x')\delta(x-y'),
\end{aligned}
\label{eq:tildeF4Gaussianstat}
\end{equation}
respectively. Equations~(\ref{eq:tildeF2Gaussianstat}) and (\ref{eq:tildeF4Gaussianstat}) exhibit the same structure as the two- and four-point correlation functions in Eqs.~(\ref{eq:tildeF2uniformstat}) and (\ref{eq:tildeF4uniformstat}), but are exact only for Gaussian probability density functions. When $P(\varphi)$ is not Gaussian, Eqs.~(\ref{eq:tildeF2Gaussianstat}) and (\ref{eq:tildeF4Gaussianstat}) only represent an approximation of the exact two- and four-point correlation function. By using Eqs.~(\ref{eq:tildeF2Gaussianstat}) and (\ref{eq:tildeF4Gaussianstat}), the two- and four-point correlation functions of the field $\tilde{E}(\omega)$ in Eq.~(\ref{eq:stochpulse}) are given by
\begin{equation}
\begin{aligned}
F_2(\omega_1, \omega_2) \doteq\,& \langle\tilde{E}^*(\omega_1)\tilde{E}(\omega_2)\rangle = \int \diff x \int \diff y \,\tilde{g}^*(x)\,\tilde{g}(y)\,\tilde{F}_2(x, y)\,\tilde{u}(\omega_1-x)\,\tilde{u}(\omega_2-y)\\
\approx\,& \tilde{g}^*(\omega_1)\,\tilde{g}(\omega_2)\,\left[2\pi\,\eu^{-\langle\varphi^2\rangle} + \left(1-\eu^{-\langle\varphi^2\rangle}\right)
\varLambda\sqrt{\pi}\tau\,\eu^{-\tfrac{(\omega_1 - \omega_2)^2\tau^2}{4}}\right]
\end{aligned}
\label{eq:F2Gaussianstat}
\end{equation}
and
\begin{equation}
\begin{aligned}
&F_4(\omega_1, \omega_2, \omega_3, \omega_4) \doteq \langle\tilde{E}^*(\omega_1)\tilde{E}(\omega_2)\tilde{E}^*(\omega_3)\tilde{E}(\omega_4)\rangle \\
=\,& \int \diff x \int \diff y\int \diff x'\int \diff y'\,\tilde{g}^*(x)\,\tilde{g}(y)\,\tilde{g}^*(x')\,\tilde{g}(y')\,\tilde{F}_4(x, y, x', y')\,\tilde{u}(\omega_1-x)\,\tilde{u}(\omega_2-y)\,\tilde{u}(\omega_3-x')\,\tilde{u}(\omega_4-y')\\
\approx\,& \tilde{g}^*(\omega_1)\,\tilde{g}(\omega_2)\,\tilde{g}^*(\omega_3)\,\tilde{g}(\omega_4)\biggl\{(2\pi)^2\,\eu^{-2\langle\varphi^2\rangle} + 2\pi\,\eu^{-\langle\varphi^2\rangle}\,\left(1-\eu^{-\langle\varphi^2\rangle}\right)\,\varLambda\sqrt{\pi}\tau\,\biggl[\biggl(\eu^{-\tfrac{(\omega_1 - \omega_2)^2\tau^2}{4}}+\eu^{-\tfrac{(\omega_1 - \omega_4)^2\tau^2}{4}}\\
&\ \ \ \ \ \ +\eu^{-\tfrac{(\omega_2 - \omega_3)^2\tau^2}{4}}+\eu^{-\tfrac{(\omega_3 - \omega_4)^2\tau^2}{4}}\biggr) - \eu^{-\langle\varphi^2\rangle} \,\biggl(\eu^{-\tfrac{(\omega_1 - \omega_3)^2\tau^2}{4}}+\eu^{-\tfrac{(\omega_2 - \omega_4)^2\tau^2}{4}}\biggr)\biggr] \\
&\ \ \ \ \ \ \ \ \ \ \ \ + \left(1-\eu^{-\langle\varphi^2\rangle}\right)^2\varLambda^2\pi\tau^2\Bigl(\eu^{-\tfrac{(\omega_1-\omega_2)^2\tau^2}{4}}\,\eu^{-\tfrac{(\omega_3-\omega_4)^2\tau^2}{4}} + \eu^{-\tfrac{(\omega_1-\omega_4)^2\tau^2}{4}}\,\eu^{-\tfrac{(\omega_2-\omega_3)^2\tau^2}{4}}\Bigr) + \cdots \biggr\},
\end{aligned}
\label{eq:F4Gaussianstat}
\end{equation}
where we have assumed a broadband frequency envelope $\tilde{g}(\omega)$. Inserted into Eq.~(\ref{eq:correlationdefinition}), the two- and four-point correlation functions in Eqs.~(\ref{eq:F2Gaussianstat}) and (\ref{eq:F4Gaussianstat}) lead to a correlation function to first order in $(\varLambda\tau)$ given by
\begin{equation}
\begin{aligned}
C(\omega_{\mathrm{R}} ,\omega_{\mathrm{m}} , T) \,&=4\pi\,\eu^{-\langle\varphi^2\rangle}\left(1-\eu^{-\langle\varphi^2\rangle}\right)^2\varLambda\Imm\biggl\{ \Bigl|\tilde{g}\Bigl(\omega_{\mathrm{m}}-\frac{\omega_{\mathrm{R}}}{2}\Bigr)\Bigr|^2\,\tilde{g}^*\Bigl(\omega_{\mathrm{m}}+\frac{\omega_{\mathrm{R}}}{2}\Bigr)\\
&\ \ \ \times \int\frac{ \diff\omega}{2\pi} \,\tilde{g}\Bigl(\omega_{\mathrm{m}}+\frac{\omega_{\mathrm{R}}}{2}-\omega\Bigr)\,\sqrt{\pi}\tau\,\biggl(\eu^{-\tfrac{\omega_{\mathrm{R}2}^2\tau^2}{4}} + \eu^{-\tfrac{(\omega_{\mathrm{R}}- \omega)^2\tau^2}{4}}\biggr)\,\eu^{-\uimm\omega T}\,\langle\hat{\tilde{\alpha}}(\omega)\rangle \biggr\}.
\end{aligned}
\end{equation}
Since $\eu^{-\langle\varphi^2\rangle}$ never vanishes, the above correlation function features the same structure as the s-TRUECARS signal in Eq.~(\ref{eq:correlationfunction}) for UDSP pulses with $a<\pi$, independent of the value of $\langle\varphi^2\rangle$.
\end{widetext}

\end{document}